\documentclass[10pt,letterpaper,twocolumn,american,aps,manuscript]{revtex4}
\usepackage[T1]{fontenc}
\usepackage[latin1]{inputenc}
\usepackage{amsmath}
\usepackage{babel}
\usepackage{graphics}

\makeatletter

\providecommand{\LyX}{L\kern-.1667em\lower.25em\hbox{Y}\kern-.125emX\@}
\newcommand{\noun}[1]{\textsc{#1}}

\makeatother

\begin{document}

\title{Jets, Stickiness and Anomalous Transport }

\author{Xavier Leoncini}

\affiliation{{\small Courant Institute of Mathematical Sciences, New York University, 251
Mercer St., New York, NY 10012, USA}\small }

\email{leoncini@cims.nyu.edu}

\author{George M. Zaslavsky}

\affiliation{{\small Courant Institute of Mathematical Sciences, New York University, 251
Mercer St., New York, NY 10012, USA }\small }

\email{zaslav@cims.nyu.edu}

\affiliation{{\small Department of Physics, New York University, 2-4 Washington Place, New
York, NY 10003, USA }\small }

\begin{abstract}
Dynamical and statistical properties of the vortex and passive particle advection
in chaotic flows generated by four and sixteen point vortices are investigated.
General transport properties of these flows are found anomalous and exhibit
a superdiffusive behavior with typical second moment exponent \( \mu \sim 1.75 \).
The origin of this anomaly is traced back to the presence of coherent structures
within the flow, the vortex cores and the region far from where vortices are
located. In the vicinity of these regions stickiness is observed and the motion
of tracers is quasi-ballistic. The chaotic nature of the underlying flow dictates
the choice for thorough analysis of transport properties. Passive tracer motion
is analyzed by measuring the mutual relative evolution of two nearby tracers.
Some tracers travel in each other vicinity for relatively large times. This
is related to an hidden order for the tracers which we call jets. Jets are localized
and found in sticky regions. Their structure is analyzed and found to be formed
of a nested sets of jets within jets. The analysis of the jet trapping time
statistics shows a quantitative agreement with the observed transport exponent.
\end{abstract}
\maketitle

\section{Introduction}

Transport phenomena can vary from electrons in conducting materials, pollutants
in the oceans or atmosphere or even data across the Internet. Typically these
phenomena are more often dealing with the transport of macroscopic scalar quantities
such as temperature or density, in other words systems for which the access
to actual microscopic information is beyond reasonable means and in some regards
overwhelming if not useless. One of the first major steps towards a proper description
of transport arose with the introduction of Fick's and Fourier's laws which
describe respectively the evolution of the density or heat current. Assuming
further simplifications both of these laws lead to the well known heat equation
and the related diffusion coefficient. The introduction of the notion of Brownian
motion and its associated probabilistic description allowed to link back this
heat equation to the microscopic world, which then is pictured as a collection
of random walkers. This may be a rather crude and oversimplified picture of
the current problems related to transport, but still today most of this probabilistic
spirit remains and in this sense the assumption of some underlying randomness
is often made. On the other hand, when considering dynamical systems, the ``microscopic''
quantities are completely or almost completely deterministic and typically evolve
with time in a ballistic or accelerated way. As a result there is a strong apparent
dichotomy underlying the diffusive or ballistic nature of transport. This dichotomy
being directly related to the properties of the underlying dynamics, and in
a sense to whether or not the dynamics preserve or loose memory. This diffusive
or ballistic nature of transport for a given system is usually inferred by the
time evolution of the second moment of its characteristic distribution, namely
\( \sim t \) for a diffusive regime and \( \sim t^{2} \) for the ballistic
one. Nature is however not so reductive and for numerous systems the behavior
\( \sim t^{\mu } \) with \( 0<\mu <2 \) or even more complicate, is observed:
transport is so-called anomalous. These anomalous properties result from a subtle
interplay of both the diffusive and ballistic behaviors and are linked to Levy-type
processes and their generalizations \cite{Montroll84}-\cite{Kovalyov2000}.

In this paper we address the question of the motion of a passive tracer evolving
in an unsteady incompressible two dimensional flow. The underlying problem is
related to the transport and mixing in fluids, or more specifically, geophysical
flows \cite{provenzale99}-\cite{Meleshko93}. In order to tackle this problem
and especially the anomalous features often observed in geophysical flows, our
approach has been gradual and the present work follows from a series of papers
\cite{KZ98}-\cite{Laforgia01}, which consists of progressive successive steps
of the investigations of problems of transport in two-dimensional flows from
the dynamical point of view. The approach originates from the uncovering of
the phenomenon of chaotic advection \cite{Aref84}-\cite{Crisanti92}, that
describes the possible chaotic nature of Lagrangian trajectories in a non chaotic
velocity field and hence reflects a non-intuitive interplay between the Eulerian
and Lagrangian perspective. The rise of chaos in these low dimensional systems
allows to considerably enhance the mixing properties which would have else to
rely on molecular diffusion. However the non-uniformity of the phase space and
the presence of islands of regular motion within the stochastic sea has a considerable
impact on the transport properties of such systems. The phenomenon of stickiness
on the boundaries of the islands generates strong ``memory effects'' as a
result of which transport becomes anomalous. In this case the rise of anomalous
transport can be directly understood by the underlying dynamics, and gives the
means of a well defined probabilistic description \cite{LKZpreprint}. However,
typical geophysical flows can not in general be considered as low dimensional
systems, hence one is tempted consider the method of two-dimensional turbulence
(i.e as high-dimensional system from the dynamical stand point) by introducing
some noise term in order to simplify the dynamics of tracers and obtain different
properties of this ad hoc noise by comparing analytical estimates with experimental
or numerical results. 

The present approach tackles these problems from another perspective, namely
a relatively simple model is chosen and a thorough analysis of the dynamics
of tracers is performed. In other words, instead of introducing noise, we mask
our ignorance by simplifying the actual system from the start and take a pure
dynamical perspective of the problem. We believe that we may in this way shed
some light on the kinetics which actually govern transport and hence complement
the more traditional probabilistic description. To settle for a model we emphasize
on another peculiarity of two-dimensional turbulent flows, namely the presence
of the inverse energy cascade, which results in the emergence of coherent vortices,
dominating the flow dynamics \cite{Benzi86}-\cite{Carnevale91}. For these
systems point vortices have been used with some success to approximate the dynamics
of finite-sized vortices \cite{Zabusky82}-\cite{VFuentes96}, as for instance
in punctuated Hamiltonian models \cite{Carnevale91,Benzi92,Weiss99}. Moreover,
point vortices have been recently used to describe exact unstationnary two-dimensional
solution of the Navier-Stockes equation \cite{Agullo01}, we may thus also envision
that the chaotic motion of the vortices shall reproduce to some extent the properties
of a more realistic flow. It therefore seemed natural to consider a system of
point vortices as our paradigm.

In the previously mentioned work, the advection in systems of three and four
point vortices evolving on the plane has been extensively investigated \cite{KZ98}-\cite{LKZpreprint}.
Three point-vortex systems on the plane have the advantage of being an integrable
system and often generate periodic flows (in co-rotating reference frame)\cite{Aref79}-\cite{Tavantzis88}.
This last property allows the use of Poincar\'e maps to investigate the phase
space of passive tracers whose motion belongs to the class of Hamiltonian systems
of \( 3/2 \) degree of freedom. A well-defined stochastic sea filled with various
islands of regular motion is observed and among these are special islands also
known as ``vortex cores'' surrounding each of the three vortices. Transport
in these systems is found to be anomalous, and the exponent characterizing the
second moment exhibit a universal value close to 3/2, in agreement with an analysis
involving fractional kinetics \cite{KZ2000,LKZpreprint}. In this system, the
origin of the anomalous properties and its multi-fractal nature is clearly linked
to the existence of islands within the stochastic sea and the phenomenon of
stickiness observed around them \cite{KZ2000,LKZpreprint}. The motion of \( N \)
point vortices on the plane is generically chaotic for \( N\ge 4 \) \cite{Novikov78}-\cite{Ziglin80}.
The periodicity is then lost when considering a system of four vortices or more,
but snapshots of the system have revealed the cores surrounding vortices are
a robust feature \cite{Babiano94}-\cite{Boatto99}, the actual accessible phase
space is in this sense non uniform and stickiness around these cores has been
observed \cite{Laforgia01}. In order to find out if these properties remain
for a large number of vortices, as well as if they may be at the origin of anomalous
features of the transport properties of these systems, a thorough analysis is
required.

In the following we investigate the advection properties of passive tracers
in flows generated by respectively 4 and 16 identical vortices. In Section II,
we recall briefly the dynamics of point vortices and of passive tracers. In
Section III the dynamics of the system of 16 vortices is investigated, basic
properties such as the time-averaged spatial distribution or minimal inter-vortex
distance are computed numerically. We observe and describe a formation of pair
and triplet of vortices, obtain statistics of pairing times in a power law tail,
implying finite average of pairing times as well as strong non trivial memory
effects. The measured pairing-time distribution exponent proves to be close
to its proposed analytical estimate. In Section IV, we consider the motion of
a passive tracer in a four vortex system and develop a new methodology for studying
the relative evolution of two nearby ``sticky'' tracers using a notion of
chaotic jet \cite{Afanasiev91}. The distribution of trapping times within jets
and the associated Lyapunov exponents are computed. The former exhibit a power-law
tail. Chaotic jets are located and are directly linked to the sticking behavior
of tracers, moreover their structure is analyzed and exhibit a nested set of
jets within jets. The introduction of a ``geometric'' Lyapunov exponent allows
to characterize independently each sticky zone. The method is then successfully
applied to the system of sixteen vortices, leading to a possible dynamical mechanism
of detecting coherent structures. In Section V, we consider transport properties
of the 16 vortices as well as those of the tracers in both systems of 4 and
16 vortices. All are found to be anomalous with characteristic exponent \( \mu \sim 1.8 \),
in good agreement with observed trapping times exponent and the kinetic theory
discussed in \cite{LKZpreprint}.

\section{Vortex and Passive Tracer Dynamics}

System of point vortices are exact solutions of the two-dimensional Euler equation
\begin{eqnarray}
\frac{\partial \Omega }{\partial t}+[\Omega ,\Psi ] & = & 0\label{Euler} \\
\Delta \Psi  & = & \Omega \: ,\label{Poisson} 
\end{eqnarray}
where \( \Omega  \) is the vorticity and \( \Psi  \) is the stream function.
The vortices describe the dynamics of the singular distribution of vorticity
\begin{equation}
\label{vorticity}
\Omega (z,t)=\sum _{\alpha =1}^{N}k_{\alpha }\delta \left( z-z_{\alpha }(t)\right) ,
\end{equation}
 where \( z \) locates a position in the complex plane, \( z_{\alpha }=x_{\alpha }+iy_{\alpha } \)
is the complex coordinate of the vortex \( \alpha  \), and \( k_{\alpha } \)
is its strength, in an ideal incompressible two-dimensional fluid. This system
can be described by a stream function acting as a Hamiltonian of a system of
\( N \) interacting particles (see for instance \cite{Lamb45}), referred to
as a system of \( N \) point vortices. The system's evolution writes
\begin{equation}
\label{vortex.eq}
k_{\alpha }\dot{z}_{\alpha }=-2i\frac{\partial H}{\partial \bar{z}_{\alpha }}\: ,\hspace {10mm}\dot{\bar{z}}_{\alpha }=2i\frac{\partial H}{\partial (k_{\alpha }z_{\alpha })}\: ,(\alpha =1,\cdots ,N)\: ,
\end{equation}
 where the couple \( (k_{\alpha }z_{\alpha },\bar{z}_{\alpha }) \) are the
conjugate variables of the Hamiltonian \( H \). The nature of the interaction
depends on the geometry of the domain occupied by fluid. For the case of an
unbounded plane, the resulting complex velocity field \( v(z,t) \) at position
\( z \) and time \( t \) is given by the sum of the individual vortex contributions:
\begin{equation}
\label{velocity_{f}ield}
v(z,t)=\frac{1}{2\pi i}\sum _{\alpha =1}^{N}k_{\alpha }\frac{1}{\bar{z}-\bar{z}_{\alpha }(t)}.
\end{equation}
 and the Hamiltonian becomes
\begin{equation}
\label{Hamiltonianvortex}
H=-\frac{1}{2\pi }\sum _{\alpha >\beta }k_{\alpha }k_{\beta }\ln |z_{\alpha }-z_{\beta }|\equiv -\frac{1}{4\pi }\ln \Lambda \: .
\end{equation}
 The translational and rotational invariance of the Hamiltonian \( H \) provides
for the motion equations (\ref{vortex.eq}) three other conserved quantities,
besides the energy, 
\begin{equation}
\label{constantofmotion1}
Q+iP=\sum ^{N}_{\alpha =1}k_{\alpha }z_{\alpha },\hspace {1.2cm}L^{2}=\sum _{\alpha =1}^{N}k_{\alpha }|z_{\alpha }|^{2}.
\end{equation}
Among the different integrals of motion, there are three independent first integrals
in involution: \( H \), \( Q^{2}+P^{2} \) and \( L^{2} \); consequently the
motion of three vortices on the infinite plane is always integrable and chaos
arises when \( N\ge 4 \) \cite{Novikov75}.

On the other hand, the evolution of a tracer is given by the advection equation
\begin{equation}
\label{gen.adv}
\dot{z}=v(z,t)
\end{equation}
 where \( z(t) \) represent the position of the tracer at time \( t \), and
\( v(z,t) \) is the velocity field (\ref{velocity_{f}ield}). For a point vortex
system, the velocity field is given by Eq. (\ref{velocity_{f}ield}), and equation
(\ref{gen.adv}) can be rewritten in a Hamiltonian form: 
\begin{equation}
\dot{z}=-2i\frac{\partial \Psi }{\partial \bar{z}},\hspace {1.2cm}\dot{\bar{z}}=2i\frac{\partial \Psi }{\partial z}
\end{equation}
 where the stream function 
\begin{equation}
\label{stream}
\Psi (z,\bar{z},t)=-\frac{1}{2\pi }\sum _{\alpha =1}^{4}k_{\alpha }\ln |z-z_{\alpha }(t)|
\end{equation}
 acts as a Hamiltonian. The stream function depends on time through the vortex
coordinates \( z_{\alpha }(t) \), implying a non-autonomous system.

In the following we focus on systems with \( N=4 \) and \( N=16 \) vortices.
Due to chaotic nature of the evolutions we rely heavily on numerical simulations.
The trajectories of the vortices as well as those of the passive tracers are
integrated numerically using the fifth-order simplectic scheme described in
\cite{McLachlan92} and which has already been successfully used in \cite{KZ98,KZ2000,LKZpreprint,Laforgia01}.

\section{16-vortex system features, pairing properties and some statistics}

\subsection{Description of the system}

We shall start by defining the system of 16 vortices which we will use to generate
the flow advecting passive tracers. As we evolve from the 4 vortex system described
in \cite{Laforgia01} to 16 vortices the phase space is considerably increased
and due to the long range 
\begin{figure}[!h]
{\par\centering \resizebox*{7cm}{!}{\includegraphics{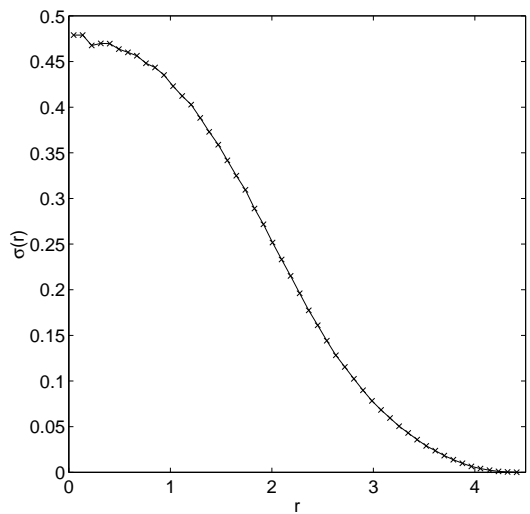}} \par}

\caption{Spatial density distribution of the vortices obtained with one trajectory computed
up to \protect\( t=10^{5}\protect \) (corresponding to \protect\( 1.6\: 10^{6}\protect \)
data points) . Due to vortex indifferenciation (permutation symmetry) and the
rotation invariance, the density depends only on \protect\( r\protect \), \emph{i.e}
on the distance from the center of vorticity, and it is identical for each individual
vortex with the same \protect\( r\protect \). We notice a bell shaped distribution
which is reminiscent of the Lamb-Oseen vortex. Note that the non-uniformity
does not necessarily imply non-ergodicity.\label{Fig16vortdensity}}
\end{figure}
interaction between vortices the energy does not behave as an extensive variable.
However the constant \( L^{2} \) defined in (\ref{constantofmotion1}) seems
to scale as \( r^{3}_{max} \), where \( r_{max} \) is the maximum distance
reached between vortices, provided that the origin of our systems corresponds
to \( Q+iP=0 \) and the vortex strengths are all equal and the vortices have
approximately a uniform distribution. In order to keep some coherence between
the four vortex system and the sixteen vortex one, we chose to keep the average
area occupied by each vortex approximatively constant. The switch from 4 to
16 vortices can then be thought of as increasing the domain with non-zero vorticity
while keeping the vorticity constant within the patch, in other words we choose
not to concentrate nor to dilute vorticity while increasing the number of vortices.
In this light we can write that the area occupied by the vortices is such that
\( r^{2}_{max}\sim N \), and thus \( L\sim N^{3/2} \), which leads to \( L=64 \)
for \( N=16 \) and is our choice for \( L \). The initial condition is chosen
randomly within a disk, we choose such a configuration so that there is no vortices
with close neighborhood to avoid any possible forced pairing. After that all
position are rescaled to match the condition \( L=64 \). The resulting simulation
show that the vortices are evolving within a disk of radius \( \sim 4 \) (see
Fig. \ref{Fig16vortdensity}), which corresponds to \( r^{2}_{max}/N\sim 1 \).
We recall that for the four vortex system with \( L=4 \), we have \( r^{2}_{max}/N\sim 1 \)
too , and that the expression \( L\sim N^{3/2} \) is not correct for only four
vortices.

We can notice in Fig. \ref{Fig16vortdensity}, that the time averaged spatial
distribution of the point vortices is not uniform, it has a bell shape which
can remind us of an extended vortex such as the Lamb-Oseen one. The stationary
distribution illustrated in Fig. \ref{Fig16vortdensity} is a time average and
in general it can not be associated with an extended stationary solution of
the Euler equation (\ref{Euler}). We shall not go into further details, but
this non-uniformity by concentrating vortices in the center is likely to lead
differences between the 4 and 16 vortex system, especially regarding the minimum
inter-vortex distance and the resulting core size, which, as it will be seen
later, are both be much smaller in the 16 vortex system than in the 4-vortex
one.

We now move on to vortex pairing and pairing time distributions.

\subsection{Minimum distance between vortices, vortex pairing and triplets}

\begin{figure}[!h]
{\par\centering \resizebox*{6cm}{!}{\includegraphics{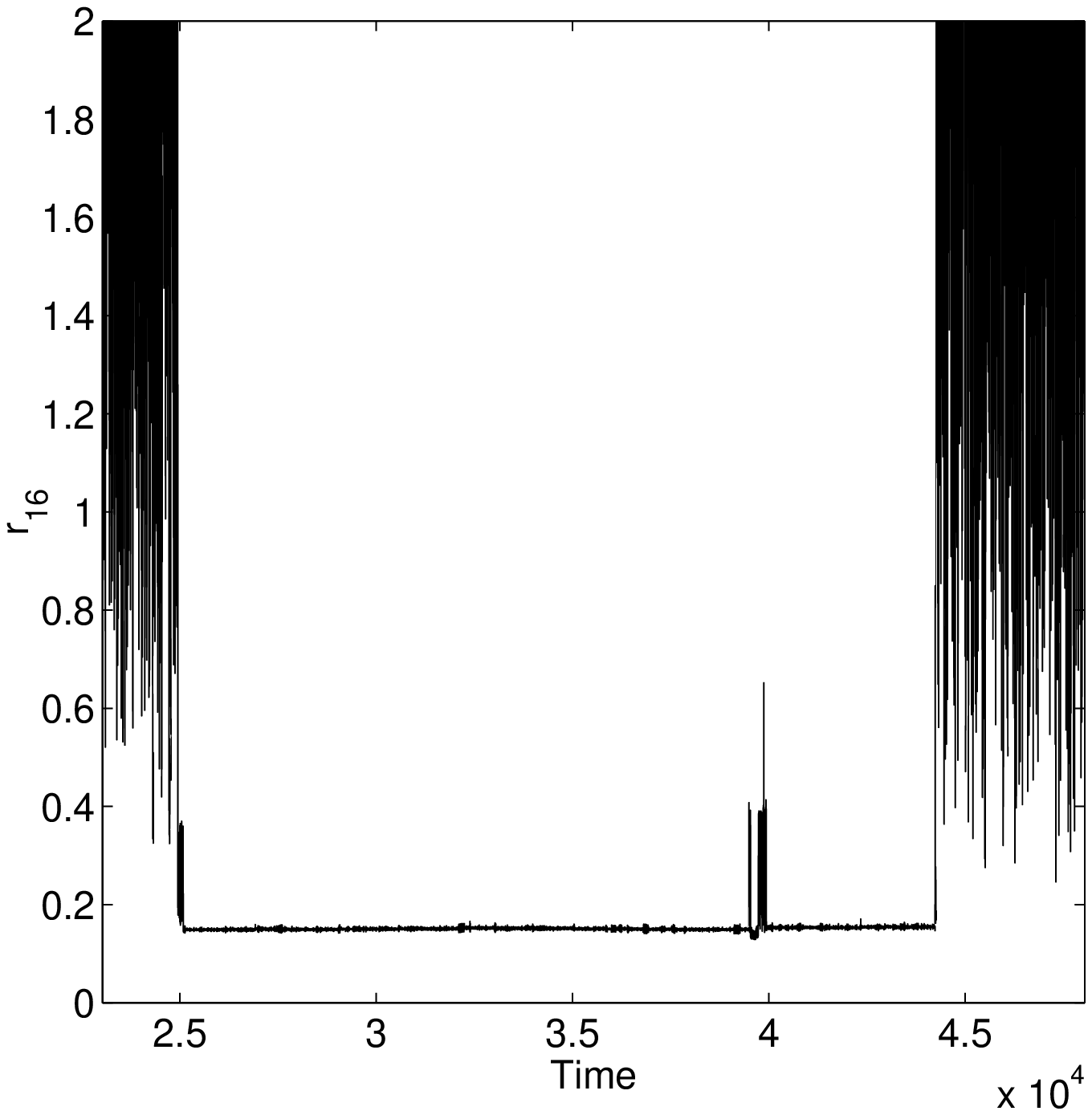}} \\
\par}

{\par\centering \resizebox*{6cm}{!}{\includegraphics{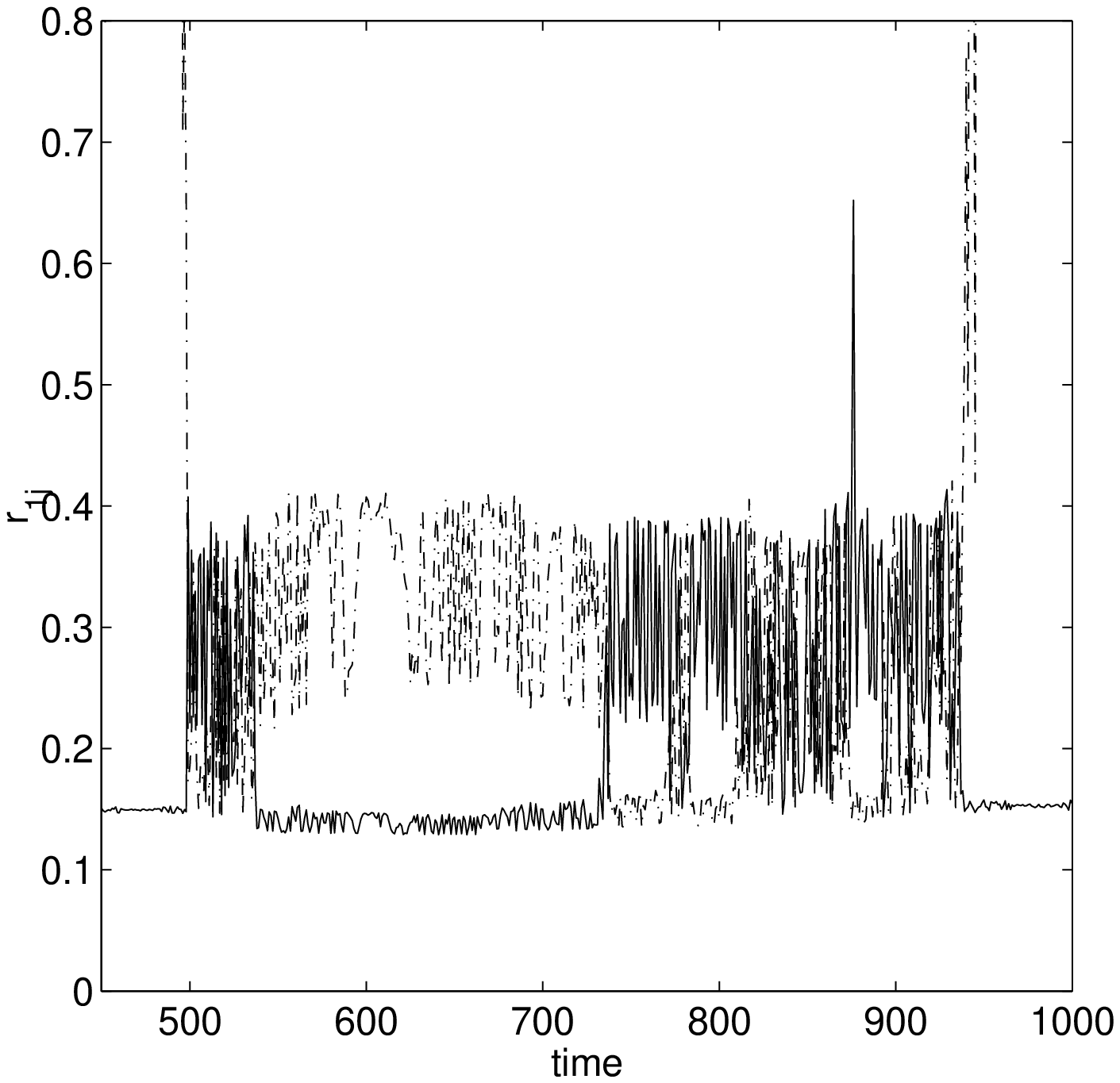}} \par}

\caption{Observation of a very long pairing of two vortices: \protect\( \Delta t\approx 2\: 10^{4}\protect \)
(upper plot). We notice a bump in the pairing around \protect\( t=4\: 10^{4}\protect \).
The analysis of this bump reveals the formation of a triplet of vortices (lower
figure) which lasts about \protect\( \Delta t\approx 450\protect \), which
is still very large compared to typical time scales. In the upper plot the relative
distance between vortex \protect\( 1\protect \) and \protect\( 6\protect \)
is plotted versus time, while in the lower plot we added also the relative distance
between vortex \protect\( 1\protect \) and \protect\( 16\protect \) (dashed
line) for the time-length of the observed bump.\label{FIr16vpairingandtriple}}
\end{figure}
\begin{figure}[!h]
{\par\centering \resizebox*{6cm}{!}{\includegraphics{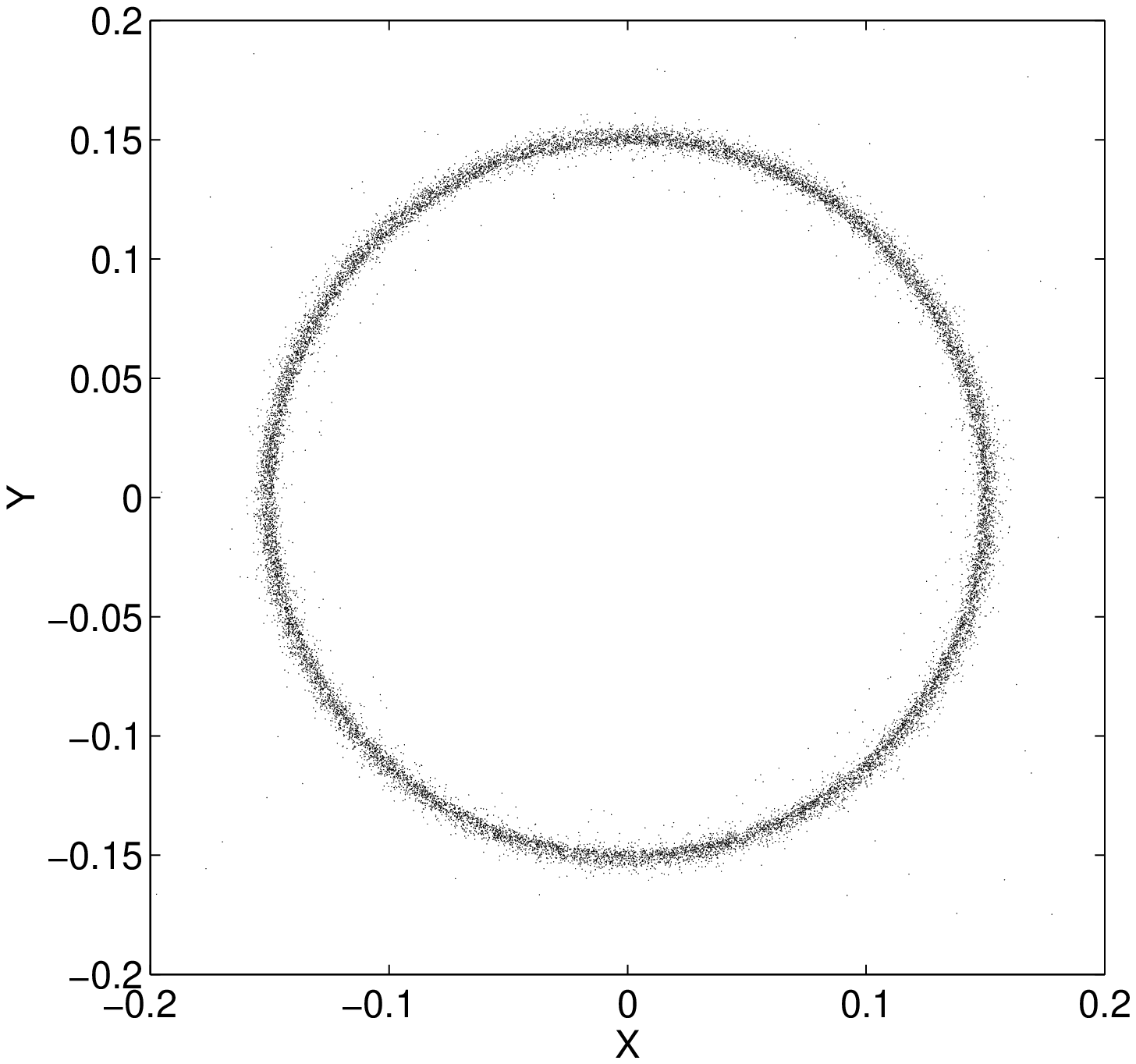}} \\
\par}

{\par\centering \resizebox*{6cm}{!}{\includegraphics{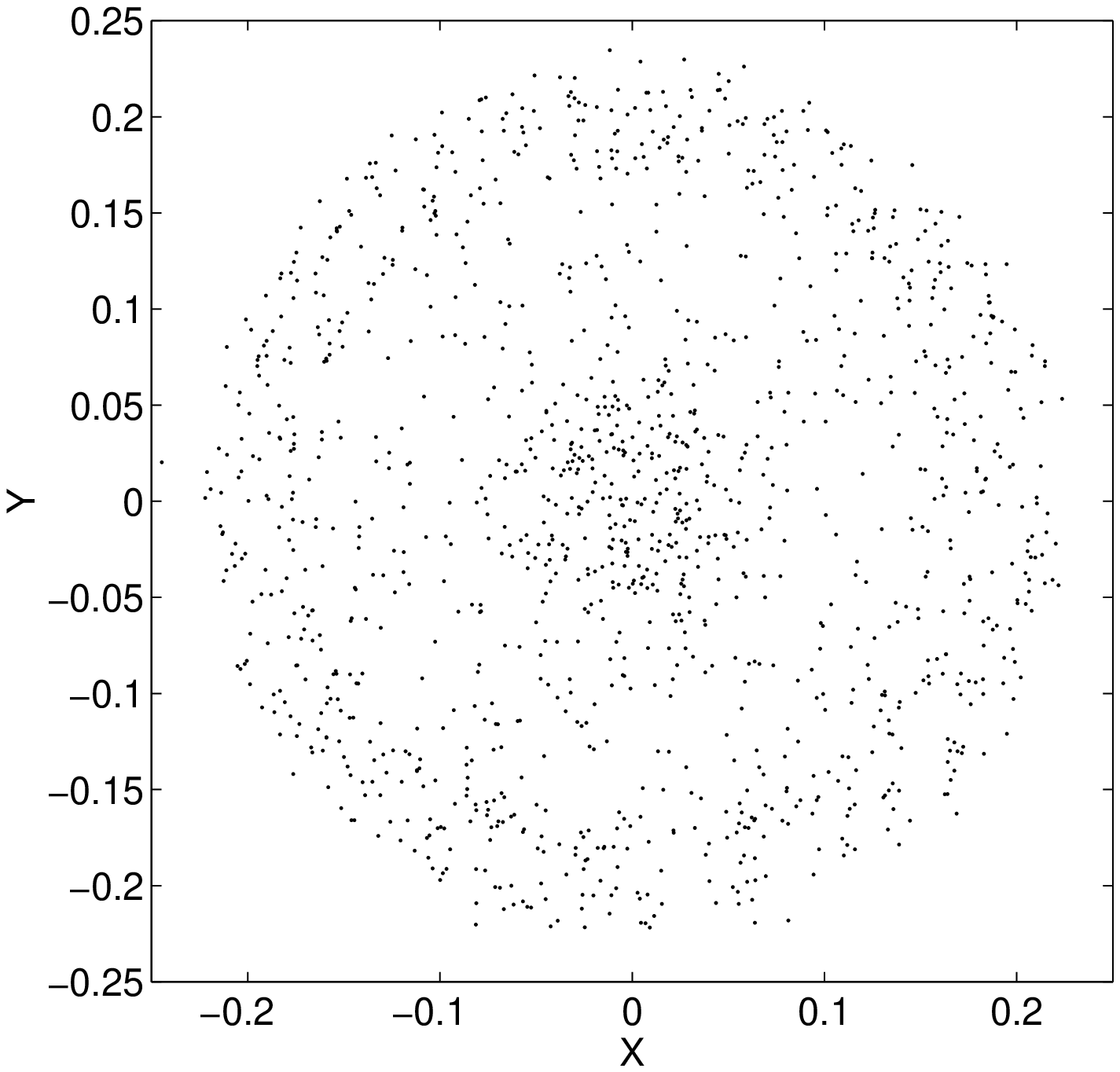}} \par}

\caption{In this figure the relative positions of the vortices involved in the pairing,
corresponding to Fig. \ref{FIr16vpairingandtriple}, are plot. The upper plot
shows the position of vortex \protect\( 6\protect \) relative to vortex \protect\( 1\protect \)
versus time. The lower plot shows the positions of vortex \protect\( 1,\: 6,\: 16\protect \)
relative to their center of vorticity. In both cases we notice that the space
occupied by the system has a typical radius of around \protect\( 0.2\protect \)
which is to be compared to an average area occupied by each vortices of \protect\( \sim 1\protect \)
. \label{Figpairandtrippos}}
\end{figure}

 It has been found in \cite{Laforgia01} for a 4-vortex system that the pairing
of vortices dramatically influences the trapping of tracers at the periphery
of the vortex cores. Namely the pairing allows the sticky region around the
cores to exchange trapped tracers, while ``opening the door'' for new tracers
to be trapped or some to escape. Since the same behavior should occur with sixteen
vortices, we decided to investigate the pairing behavior of the considered \( 16 \)-vortex
system. For this purpose we carried out a simulation up to \( t=10^{5} \),
and checked the behavior of inter-vortex distances versus time. The results
indicate that long time vortex pairing exists and one vortex pairing which lasts
\( \Delta t\sim 10^{4} \) is illustrated in Fig. \ref{FIr16vpairingandtriple}.
We also notice that during the pairing a triplet (a system of \( 3 \) bound
vortices) is formed for about \( \Delta t\sim 500 \). The phenomenon of formations
of triplets and pair of vortices concentrates vorticity in small regions of
the plane (see Fig. \ref{Figpairandtrippos}) and in some sense is reminiscent
of the inverse energy cascade observed in 2D turbulence. 
Since no quadruples are observed, we recall that both systems of 2 or 3 vortices
are integrable, and may hence wonder whether the observation of triplets and
pairs is just pure coincidence or that the long memory effects associated with
stickiness are intimately linked to this type of behavior. Namely, for passive
tracers in the three vortex systems, the phenomenon of stickiness is associated
with tracers that stay a ``long time'' in the vicinity of a island and mimic
the regular trajectory of tracers trapped within the island. This notion was
somewhat extended in \cite{Laforgia01}, where the pairing behavior in the 4-vortex
system was described as a sticking phenomenon to an object of lesser dimension
than the whole phase space. However we speak about the pairing in the 4-vortex
system as a reduction to an integrable 3-vortex system. It is therefore tempting
to generalize this behavior as a sticking phenomenon to an object of lesser
dimension than the actual phase space, but with some constraints. The subspace
is reached by generating subsystems which integrability is a good approximate
for a fairly long time. In this light, stickiness would put some condition on
the actual structure of potential clusters of vortices, for instance quadruples
would be possible to encounter only if two out of the four vortices are involved
in a pairing on a smaller scale, giving rise to a triplet on a larger scale.
However we have not confirmed nor infirmed this scenario with the system of
16 vortices, and hence shall keep for now the simpler generalized notion of
stickiness defined in \cite{Ott98,ZEN97,Laforgia01}. In any case, both the
triplet and pairing events described in Fig. \ref{FIr16vpairingandtriple} correspond
to a sticking behavior, as the system remains a long time on a given subset
of the phase space. For comparison we mention that a typical time of an eddy
turnover used in \cite{Weiss98}, corresponds to a time of order \( \Delta t\sim 1-5 \).

Finally by detecting the pairing of vortices we were, at the same time, able
to measure the minimum distance between vortices. It was suggested in \cite{Babiano94}
that the minimum distance can be a pretty good indicator of the double of the
size of the vortex cores surrounding the vortices \cite{LKZpreprint,Laforgia01}.
We found out that \( \min (r_{ij})\approx 0.13 \) which implies that the core
radius estimate \( r \), given by a half of the minimum of the inter-vortex
distance should be of the order \( r\sim 0.065 \).

\subsection{Paring time statistics}

\begin{figure}[!h]
{\par\centering \resizebox*{7cm}{!}{\includegraphics{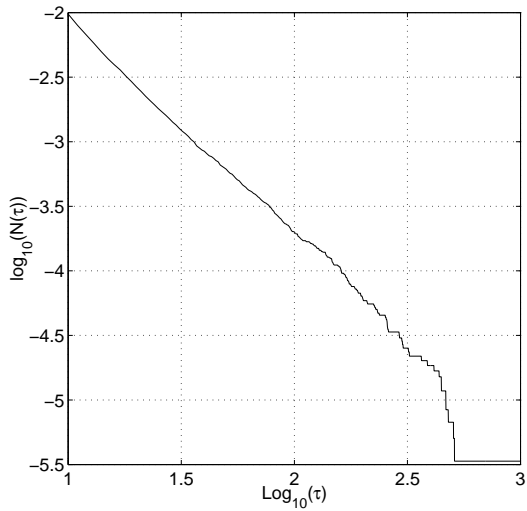}} \par}

\caption{Distribution of integrated pairing times (see definition (\ref{probdef}))
for the system of 16 vortices. We notice an algebraic time decay \protect\( N\sim \tau ^{-\gamma _{p}+1}\protect \)
with \protect\( \gamma _{p}\approx 2.68\protect \).\label{Figpariingtimedist}}
\end{figure}

In the previously mentioned works \cite{KZ2000,LKZpreprint}, it has been shown
that the stickiness, providing long coherent motion, leads to anomalous transport
properties and distributions with power law tails. We will show how the vortex
pairing is related to the stickiness phenomenon and how it influences on the
motion of passive tracers. Following the methodology and the results presented
for 4 vortices in \cite{Laforgia01}, we consider statistical data on pairing
times for the 16-vortex system, using the previously run simulation of vortex
motion up to time \( t=10^{5} \). The detection of pairing events is obtained
with the same technique directly inspired from Fig. \ref{FIr16vpairingandtriple}:
a pairing occurs if for a given length of time two vortices stay close together.
The results obtained for the four vortex system were independent on the arbitrary
cutoffs chosen to characterize a pairing event, hence we chose the arbitrary
time length to be \( \delta t=5 \) (this value does not affect the behavior
of large pairing time), and the distance from one vortex to another is such
that \( r_{ij}=|z_{i}-z_{j}|\le 1 \). To gather the statistics we proceed as
was done in \cite{Laforgia01} and computed the the integrated probability \( N(\tau ) \)
of pairings that last longer than a time \( \tau  \)
\begin{equation}
\label{probdef}
N(\tau )=N(T>\tau )\sim \int ^{\infty }_{\tau }\rho (T)dT\: ,
\end{equation}
where \( \rho (T)dT \) is the probability density of an event to last a time
\( T \). The results are shown in Fig. \ref{Figpariingtimedist}. The analysis
of the distribution tail gives a power-law decay of \( N(\tau )\sim \tau ^{-\gamma _{p}+1} \)
with the pairing exponent \( \gamma _{p}\sim 2.68\pm 0.1 \) that confirms the
non-negligible occurrence of long lasting pairings. The behavior of the probability
density of pairing \( \rho (\tau ) \) lasting a time \( \tau  \), is obtained
from Eq. (\ref{probdef})
\begin{equation}
\label{rhobeahvior}
\rho (\tau )\sim \frac{dN}{d\tau }\sim \frac{1}{t^{\gamma _{p}}}\: .
\end{equation}
 This behavior provides a finite mean pairing time, but the second moment is
infinite if the value of \( \gamma _{p} \) can be extrapolated. As suggested
in the next section, pairing leads to one of the many form of stickiness, hence
the pairing-times are to link to the trapping time within a sticky domain of
phase space and parameters. 

In the following subsection, we provide an estimate of \( \gamma _{p} \) .

\subsection{Pairing exponent}

The main idea used to estimate the value of the paring exponent \( \gamma _{p} \)
follows the results presented in Refs.\cite{Zaslavsky2000} and \cite{LKZpreprint}.
The idea revolves around the presence of an island of stability leading to ballistic
or accelerator modes within the island. These islands appear in the stochastic
sea as a result of a parabolic bifurcation \cite{Melnikov96} and correspond
to the so-called tangled islands \cite{romkedar99}. This is fairly general
and it is reasonable to link the sticky phenomenon of vortex-pairing to the
rise of an island in the stochastic sea, \emph{i.e} the formation of virtual
potential well for the dynamics of a pair of vortices. Another way to validate
this point of view comes from the pair perspective. While the pair exists an
integrable system is formed which is perturbed by the flow generated by other
vortices. To deal with the problem we use the general form of effective Hamiltonian
proposed in \cite{Melnikov96} (see also \cite{Zaslavsky2000}and \cite{ZEN97}):
\begin{equation}
\label{Heffecpair}
H_{eff}=b(\Delta P)^{2}-a\Delta Q-V_{3}(\Delta Q)\: ,
\end{equation}
where \( \Delta P \) and \( \Delta Q \) are respectively generalized momentum
and generalized coordinate of the pair of vortices, the interaction potential
\( V_{3} \) is a third order polynomial and \( a \), \( b \) are constants.
The higher order terms in \( \Delta Q \) can be neglected for the effective
Hamiltonian. 

Let us assume that the pairing corresponds to the occurrence of an island in
the stochastic sea and that effective regular trajectories of the pair can be
described by \( H_{eff} \) given in (\ref{Heffecpair}). This island has an
elliptic point \( \xi _{e}=(P_{e},Q_{e}) \). Since the island has a finite
size, a typical trajectory \( \xi =(P,Q) \), located within the island corresponds
to periodic or quasiperiodic dynamics and can be characterized by its relative
coordinates \( (\Delta Q,\Delta P)=\xi -\xi _{e} \). When the boundary of the
island is reached, the values of the generalized coordinates \( \xi ^{*}=(P^{*}(t),Q^{*}(t)) \)
are such that the trajectory can access the whole stochastic sea, but cannot
enter the island (the generalized phase space is two-dimensional). 

The following steps are fairly formal (see also \cite{ZEN97} and \cite{LKZpreprint}).
Let us consider a trajectory which is close to the island's edge, which we monitor
by the coordinates \( (\delta P,\delta Q)=\xi -\xi * \). A small perturbation
is then likely to allow the trajectory to ``escape'' from the island vicinity
and consequently to destroy the vortex pair. The phase volume of the escaping
trajectory writes:
\begin{equation}
\label{phasevolume}
\delta \Gamma =\delta P\delta Q\: ,
\end{equation}
where \( \delta P \) , \( \delta Q \) are the values of the escaping trajectory.
Since the trajectory is close to the islands edge (inflexion point), we can
estimate from Eq.(\ref{Heffecpair})
\begin{equation}
\label{deltaPmax}
\delta P\sim \delta Q^{\frac{3}{2}},
\end{equation}
 where we have assumed \( V_{3}(Q)\sim Q^{3} \). Using this last expression
(\ref{deltaPmax}), we obtain for (\ref{phasevolume})
\begin{equation}
\label{phasevolumebis}
\delta \Gamma =\delta Q^{\frac{3}{2}}\delta Q\sim \delta Q^{\frac{5}{2}}\: .
\end{equation}
 Due to the periodic or quasiperiodic nature of the trajectories within the
island, any sticking trajectory (in its neighborhood) experiences a ballistic
type behavior, which translates into \( \delta Q\sim t \), \emph{i.e} 
\begin{equation}
\label{phasevolumeter}
\delta \Gamma \sim t^{\frac{5}{2}}\: .
\end{equation}
The probability density to escape the island vicinity after being in its neighborhood
for a time \( t \) (\emph{i.e} time-length of the pairing) within an interval
\( dt \) is then
\begin{equation}
\label{probdistbis}
\rho (t)\propto \frac{1}{\delta \Gamma (t)}\sim t^{-\frac{5}{2}}\: ,
\end{equation}
 this results gives us directly the estimate of the exponent \( \gamma _{p}\approx 5/2 \),
which is very close to the observed value \( 2.7\pm 0.1 \).

This estimate is not rigorous and essentially based on phenomenological grounds,
however we believe it provides a good insight on the origin of different characteristic
exponents of trapping time distributions. We now remind the reader that for
the 4-vortex system the value observed in \cite{Laforgia01} for characteristic
exponent was \( \gamma _{p}\approx 7/2 \). This value was explained in a very
similar way as what has just been developed before but one more generalized
spatial coordinates was introduced. A 4-vortex system is non generic. Indeed
as a pair is formed, the whole system becomes a quasi 3-vortex system, the pair
acting as one vortex with increased strength, hence the remaining system is
itself integrable. In this light, when switching to the idealized generalized
variables \( (P,Q) \), we should consider more degrees of freedom to describe
the pair of vortices, namely
\begin{equation}
\label{Heffective4vort}
H_{eff}=c_{1}(\Delta P_{1})^{2}+c_{2}(\Delta P_{2})^{2}+V_{3}(\Delta Q_{1},\Delta Q_{2})\: .
\end{equation}
The Hamiltonian (\ref{Heffective4vort}) is in general non-integral, and the
appearance of an island of stability adds an additional constraint or integral
of motion to the system governed by (\ref{Heffective4vort}). Taking this constraint
into account the Hamiltonian (\ref{Heffective4vort}) can be transformed into
\begin{equation}
\label{Hamilteffect4bis}
H_{eff}=c(\Delta P)^{2}+V_{3}(\Delta Q_{1},\Delta Q_{2})\: ,
\end{equation}
where \( \Delta P \) is a new (collective) momentum. The corresponding phase
volume of the escaping trajectories gives in analogy to (\ref{phasevolume}):
\begin{equation}
\label{phasevol4vort}
\delta \Gamma =\delta P\delta Q_{1}\delta Q_{2}\sim \delta Q^{5/2},\hspace {10mm\}}\left( Q\sim \delta Q_{1,2}\right) 
\end{equation}
This leads to the estimate of \( \gamma _{p}\approx 7/2 \), in contrary to
Eq. (\ref{probdistbis}) with \( \gamma _{p}\approx 5/2 \). Note that the extra
spatial generalized coordinate introduced in \cite{Laforgia01}, can also be
thought as a consequence of having two different coexisting quasi-integrable
subsystem described by different actions, but however linked as they cannot
``live'' without one-another.

We have now a sufficient knowledge of the dynamics of the vortex systems, and
move on to the behavior of passive tracers generated by these flows.

\section{Jets}

\subsection{Definitions}

As it was previously mentioned, the motion of point vortices is chaotic for
both systems of four or sixteen vortices but the use of Poincare maps in these
cases is impossible in contrary to \cite{KZ98,KZ2000,LKZpreprint}. To investigate
the anomalous transport properties from the first principles it is crucial to
define a proper diagnostic which will be able to capture some singular properties
of the dynamics that are clues of the origin of the anomalous transport of passive
tracers. 

For the system of 4 point vortices, successive snapshots have shown that passive
tracers can stick on the boundaries of cores and jump from one core to another
core or escape from the core due to their perturbations. The fact that a tracer
is able to escape from a core means that the surrounding regions of the cores
are connected to the region of strong chaos. The results presented in \cite{Laforgia01}
indicate that these regions mix poorly with the region of strong chaos. One
way to track this phenomenon is to use Finite-Time Lyapunov exponents (FTLE)
and to eliminate domains of small values of the FTLE \cite{Boatto99,Castiglione99,Andersen00}.
Once these exponents are measured from tracers' trajectories whose initial conditions
are covering the plane, a scalar field distributed within the space of initial
conditions is obtained and the two dimensional plot of the scalar field reveals
regions of vanishing FTLE, namely the cores surrounding the vortices and the
far field region. The cores are thus regions of small FTLE, meaning that two
nearby trajectories are bound together for long times, and this despite the
core's chaotic motion. These properties reveal typically a sharp change of the
tracers dynamics as it crosses from the region of strong chaos to the core.
This property is directly linked to the method described in in \cite{Haller2000},
which determines from a Lagrangian perspective the border of coherent structures
in a turbulent flow. Namely, \noun{\Large }the method consists in computing
a scalar field (typically FTLE's) and extract the coherent structures by finding
the spatial extrema of this scalar field. The difficulties with these types
of approach resides in the definition of the Lyapunov exponent 
\begin{equation}
\label{lyapunovdefinition}
\sigma _{L}=\lim _{r(0)\rightarrow 0}\lim _{\tau \rightarrow \infty }\frac{1}{\tau }\ln \frac{r(\tau )}{r_{0}}\: ,
\end{equation}
where \( r_{0} \) is the initial separation between two nearby trajectories
and \( r(\tau ) \) is the separation at time \( \tau  \). Indeed, the definition
(\ref{lyapunovdefinition}) introduces an arbitrary choice of two free parameters
when computing FTLE, namely the initial separation between two different trajectories
\( r_{0} \) and the time interval \( \tau  \) within which they are computed.
Moreover FTLE are not unique for a given trajectory which induces also a dependence
on initial conditions \( x_{0} \), as well as time if the system is not autonomous.
In the following we shall note FTLE as \( \sigma _{L} \), but shall keep in
mind that \( \sigma _{L}=\sigma _{L}(\tau ,r_{0},x_{0},t) \).

One problem that may arise when computing these exponents is related to the
behavior of \( r(t) \). For instance in the case of a system of four identical
point vortices, the motion of tracers is more or less confined within a finite
sized region and \( r(t) \) has an upper boundary \( R(t) \) which may be
dependent on time if radial diffusion occurs, and no matter what initial distance
\( r_{0} \) and initial position \( x_{0} \), \( \sigma _{L}\rightarrow 0 \)
as \( \tau \rightarrow \infty  \). This example is rather extreme and exaggerated
as the results presented in \cite{Boatto99} are able to capture the structure
of the space of initial conditions, but we believe it illustrates clearly one
problem encountered when using FTLE, namely that \( r(t) \) is dependent on
possible scales of the physical nature of the system. It is likely that \( r(t) \)
is not always smooth growing function of time on the scale of an arbitrary time
\( \tau  \) and jumps between different spatial scales with a potential physical
meaning which may get averaged with time. We can anticipate that this may be
especially the case when different regions of small (if not zero) Lyapunov exponents
are present in the system.

In the previous discussion it appears that FTLE are overall giving us good results
in detecting coherent structures and regions of low chaos, but may also hide
by averaging out a useful information as a result of a lack of capturing specific
scales. In the following we propose an alternative diagnostic which is greatly
inspired from typical FTLE but has the advantage of clearing out some of its
shortcomings. Namely, FTLE is a straightforward approximation of the definition
(\ref{lyapunovdefinition}) of the Lyapunov exponent which is is inherently
nonlocal. In other words a Lyapunov exponent measures the ``averaged'' exponential
divergence of two nearby trajectories, and assuming the system is ergodic it
measures a degree of ``chaoticity'' of the whole dynamics of the considered
system. This non-locality property may create serious difficulties in the interpretation
of the results when the truncated characteristics of the dynamics has been used
while for the considered time interval the ergodic theorem may not work (see
more discussion in {[}\cite{Afanasiev91}). 
\begin{figure}[!h]
{\par\centering \resizebox*{7cm}{!}{\includegraphics{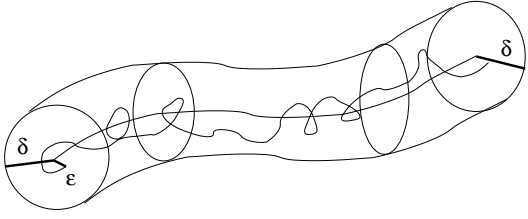}} \par}

\caption{A tracer and a ghost are used to define a jet.}
\end{figure}

One possible way to circumvent this problem can be identified by the following
remarks. We are most of the time dealing with only a finite portion of a trajectory
(finite time) and only have a finite spatial resolution of these pieces of trajectories
which is typically much smaller than the actual scales we are interested in.
In this more practical situation, we are facing a ``coarse grained'' phase
space, and each point is actually a ball from which infinitely many real trajectories
can depart. Given this facts we can imagine that two nearby real trajectories
diverge exponentially for a while but then get closer again without actually
leaving the ball, a process which may happen over and over in the case of stickiness.
From the ``coarse grained'' perspective those two real trajectories are identical.
We then can infer that there exists bunch of nearby trajectories which may remain
within the ball for a given time, giving rise to what is called a \emph{jet}
\cite{Afanasiev91}, and can be understood as a region of regular motion for
our scale of interest. We are then mostly interested on the chaotic properties
of the system from the ball scale and up and dismay any chaotic motion which
may occur within the jet. The stickiness to a randomly moving and not well determined
in phase space coherent structure imposes existence of jets, while the opposite
may not be the case.

To actually measure the jets properties of the system, we use the following
strategy. Let us consider a given trajectory \( \mathbf{r}(t) \) evolving within
the phase space. For each instant \( t \), we consider a ball \( B(\mathbf{r}(t),\delta ) \)
of radius \( \delta  \) centered on our reference trajectory. We then start
a number of trajectories within the ball at a given time, and measure the time
it actually takes them to escape the ball. Depending on how fast trajectories
are escaping the ball we should then be able to identify if the reference trajectory
is moving within a regular jet or is in a region of strong chaos.

\subsection{Statistical Results}

From the numerical point of view we first consider the velocity field generated
by the chaotic motion of four point vortices and proceed as follows: given an
initial condition of a tracer, two ``ghost'' particles are placed in its neighborhood,
at a distance \( \epsilon =10^{-6} \) , more specifically we placed one ghost
along the tracer speed and the other one on the orthogonal direction, but this
positioning should not affect the results. Then for each of the ghost particles,
once each reaches a distance \( \delta =0.03 \) (the radius of our ball) from
the tracer, the time interval \( \Delta t \) and the distance traveled \( \Delta s \)
are recorded. For simplicity two new ghost particles are placed within the ball
once both have escaped, while the old ones are discarded. One of the main difficulties
in using this type of diagnostic relies in the fact that data acquisition is
a priori not linear in time nor space and necessitates a careful choice for
the values of \( \epsilon  \) and \( \delta  \). Note that the value chosen
for \( \delta  \) is small even compared to the minimum inter-vortex distance.
However, using the definition (\ref{lyapunovdefinition}) we can compute a different
type of FTLE, which we define as follows: 
\begin{equation}
\label{lypundef}
\sigma _{L}=\frac{1}{\Delta t}\ln \frac{\delta }{\epsilon }\: ,\hspace {10mm}\sigma _{D}=\frac{1}{\Delta s}\ln \frac{\delta }{\epsilon }\: ,
\end{equation}
where contrary to the typical definitions the value of the logarithm is fixed
and \( \Delta t \) or \( \Delta s \) are the variables. These exponent are
very similar to the notion of Finite Size Lyapunov Exponent (FSLE) considered
in \cite{Aurell97}, however we do not perform averages over different scales
and keep the whole distribution function. We computed these exponents for the
flow generated by four vortices, with the same initial condition as the one
used in \cite{Laforgia01}. The data are obtained using the trajectories of
4 different tracers initially placed in the region of strong chaos and advected
by the chaotic flow generated by the motion of the 4 point vortices of equal
strength. The time of the simulation is \( 5.10^{6} \), the time step is \( 0.05 \),
which leads to statistics computed using \( \sim 3.10^{5} \) data points. 
\begin{figure}[!h]
{\par\centering \resizebox*{7cm}{!}{\includegraphics{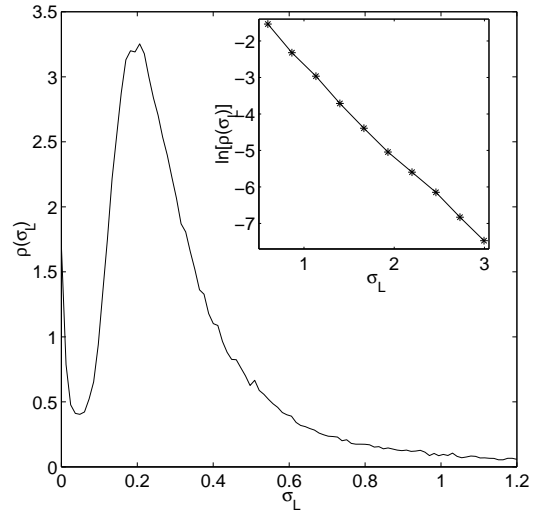}} \par}

\caption{Distribution of time Lyapunov exponents \protect\( \sigma _{L}\protect \)
(see Eq. (\ref{lypundef})). We notice an exponential decay for high exponents.
\protect\( \rho (\sigma _{L})\sim \exp (-\sigma _{L}/\sigma _{L_{0}})\protect \)
with \protect\( \sigma _{L_{0}}\approx 0.4\protect \). We can notice a minimum
around \protect\( \sigma _{L}\approx 0.05\protect \). The observed accumulation
near 0 results from the existence of long lived jets. The local minimum gives
an estimate of the minimum typical time interval corresponding to a jet: \protect\( \Delta t_{min}\approx 206\protect \).
Data are obtained with 4 different trajectories computed up to \protect\( t=5.10^{6}\protect \)
leading to 328.220 records.\label{Figlyapunovtime}}
\end{figure}
 The results of the measured \( \sigma _{L} \) are illustrated in Fig. \ref{Figlyapunovtime}.
One can notice in this plot two different types of behavior. 

First, the large FTLE decay exponentially with a characteristic exponent \( \sigma _{L_{0}}\approx 0.4 \),
which is not a surprise since the speed of tracers is bounded. Hence even if
the tracer and the ghost are going in opposite direction, it will always take
them a finite time to escape from the ball, thus an expected maximum value for
\( \sigma _{L} \). Regarding the exponential decay behavior before reaching
this maximum value, we can suspect it is directly related to the way the data
is acquired; indeed we remind the reader that the acquisition is a priori nonlinear
in time, and that we are in fact measuring escape times from a given moving
region of the phase space, and since this behavior is related to the large FTLE's,
we are just observing the exponential growth of the coarse grained volume.

\begin{figure}[!h]
{\par\centering \resizebox*{7cm}{!}{\includegraphics{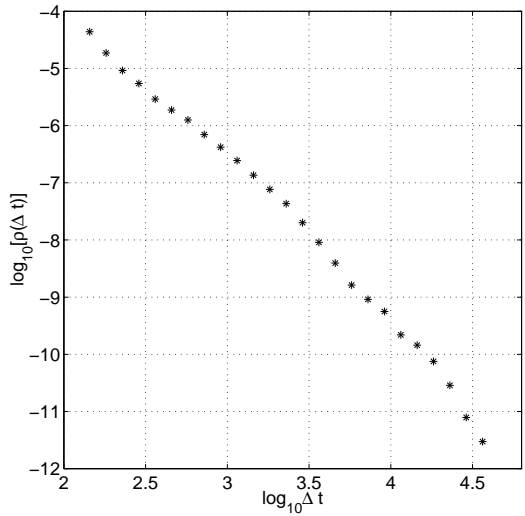}} \par}

\caption{Tail of the distribution of time intervals \protect\( \Delta t\protect \).
We notice a power-law decay, with some oscillations. Typical exponent is \protect\( \rho (t)\sim t^{-\gamma }\protect \)
with \protect\( \gamma \approx 2.823\protect \). Data is obtained with 4 different
trajectories computed up to \protect\( t=5.10^{6}\protect \) leading to 328.220
records.\label{Figescapetime4vortex}}
\end{figure}

The second behavior is, from the point of view of anomalous transport, more
interesting. The local minimum for small FTLE can be seen in the probability
density of \( \sigma _{L} \) as a crossover from the erratic chaotic motion
of tracer within the chaotic region, to its motion within a quasi-regular jet.
Indeed if the tracer is within jet, the ghosts are nevertheless expected to
escape from the tracers vicinity but with trapping times exhibiting a power-law
decay, therefore if the passive tracer is evolving within a jet for a long time,
we should expect an accumulations of events corresponding to ghosts leaving
the surrounding ball. This hypothesis is confirmed in Fig. \ref{Figescapetime4vortex},
where the distribution of trapping times is plotted. The log-log plot clearly
shows the power-law decay of the trapping times, with typical exponent \( \gamma =2.82 \). 

\begin{figure}[!h]
{\par\centering \resizebox*{7cm}{!}{\includegraphics{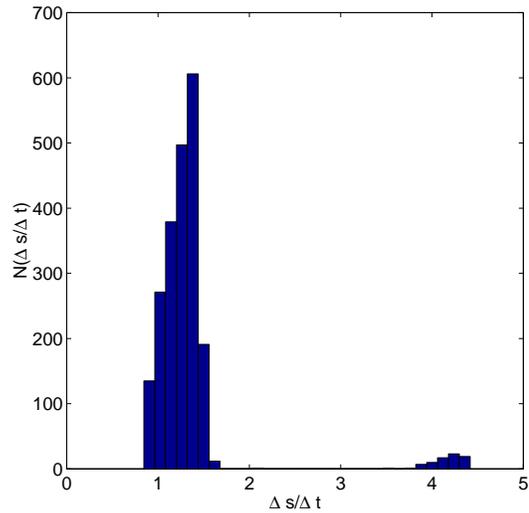}} \par}

\caption{Distribution of the averaged speed of the tracer, for the data corresponding
to \protect\( \sigma _{D}<0.03\protect \). We notice two regimes, the regime
of fast speed corresponds to stickiness to the core. Note that if instead we
use \protect\( \sigma _{L}\protect \) as a reference most of the fast particles
lie after the local minimum in the distribution and mostly only one regime seems
to be present.\label{FIgspeeddist}}
\end{figure}

We shall now discuss the reason why another Lyapunov exponent \( \sigma _{D} \)
was introduced in (\ref{lypundef}) . By its definition \( \sigma _{D} \) measures
how much two trajectories diverge depending on how much we travel along them,
it is then inherently time independent and can be seen as a pure geometric property
of a trajectory or, from another point of view, time is locally rescaled depending
on the local speed, so that the speed along the trajectory is constant and equal
to one. The plot of the distribution of \( \sigma _{D} \) gives the same picture
as the one obtained for \( \sigma _{L} \) in Fig. \ref{Figlyapunovtime}, with
an exponential decay and a a local minimum \( \sigma _{D}*\sim 0.03 \) near
zero, which also can be used as a criterion for identifying a coherent jet.
We may argue that since the speed is bounded and almost everywhere non zero,
the use of \( \sigma _{D} \) is redundant and therefore futile. Nevertheless
from a practical point of view, the interval of possible speed is rather large;
for instance in the case of the 4 vortex system, the core has a typical radius
of \( 0.2 \) while the outer region corresponds to radii of around \( 4 \),
we can thus expect an order of magnitude between the different speeds within
the region of strong chaos and the outer region, moreover we can expect an increase
of the range as we increase the number of vortices. It then becomes obvious
that by measuring \( \sigma _{L} \) we are biased towards jets occurring in
the outer region, while by using \( \sigma _{D} \) these dynamical differences
are erased and only the actual topology of the vicinity of a trajectory matters.
Hence when using \( \sigma _{D}* \) to characterize a jet in a small simulation
we obtained for the distribution of characteristic speeds the histogram plotted
in Fig. \ref{FIgspeeddist}
 , where it clearly shows that actual fast jets exists, while if we had used
only \( \sigma _{L} \) only a few fast jets are detected. In this light it
seems that \( \sigma _{D}* \) is a good candidate to identify a jet, while
its averaged speed \( \sigma _{L}/\sigma _{D} \) gives a more refined idea
on the nature of the jet.

\subsection{Jets pictures}

Given the previous results we shall now have a closer look at those jets. Namely,
in the previous sections we defined what we considered a jet, computed statistics
on them and were able with the results to obtain a threshold for which a jet
can be considered regular. We just then have to put these results in application.
Let us initialize a tracer in the region of strong chaos, but not in the vicinity
of any vortex to avoid any trapping within a core. We can let the tracer evolve
with its two ghosts nearby, once the threshold given by \( \sigma _{D}* \)
is reached (ghosts are still within the ball for a given length traveled) we
know that the measured \( \sigma _{D} \) will be such that \( \sigma _{D}<\sigma _{D}^{*} \),
hence we are currently within what we considered a regular jet, we then just
have to record the position of the tracer and vortices until ghosts have escaped.
In this way we are able to locate the tracer for given length of time while
it evolves within the jet. In fact we shall be even more picky, namely we know
from Fig. \ref{FIgspeeddist}, that the majority of jets correspond to slow
motion, which when plotted corresponds to the tracer being in the far field
region and simply rotating around the center of vorticity, hence to avoid recording
the position of these events we can also use the averaged speed \( \sigma _{L}/\sigma _{D} \)
and record only the jets corresponding to fast motion. 
\begin{figure}[!h]
{\par\centering \resizebox*{8cm}{!}{\includegraphics{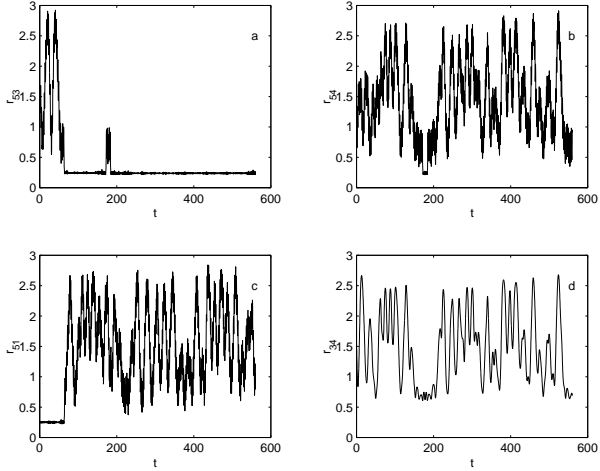}} \par}

\caption{Plots of relative distance for an identified jet within the ``strong chaos''
area whose\protect\( \Delta t\approx 560\protect \). Plots a,b,c: distances
between the passive tracer and respectively vortex 3,4,1 versus time (the tracer
is referred as particle 5). Plot d, distance between vortex 3 and 4 versus time.
We notice that for this jet the passive tracer sticks to vortex cores. It can
jump from core to core as vortex are pairing but is always sticking to one core.\label{Figreldist}}
\end{figure}
 
\begin{figure}[!h]
{\par\centering \resizebox*{8cm}{!}{\includegraphics{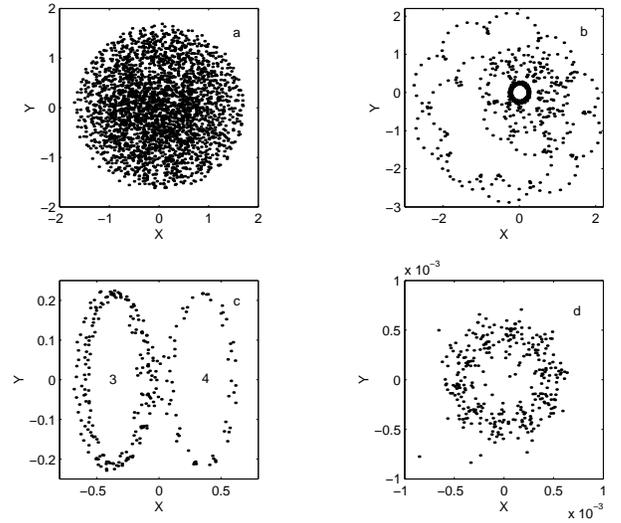}} \par}

\caption{Plots of tracer's positions for different times for an identified jet within
the ``strong chaos'' area whose\protect\( \Delta t\approx 560\protect \).
Plot a, position in the absolute reference frame (it looks random!). Plot b,
relative position of the tracer with respect to vortex 3 (see Fig. \ref{Figreldist}),
we observe the sticking habits of the tracer. Plot c, relative position of the
tracer with respect to the pair formed by vortex 3 and 4, during the pairing
(\protect\( 153<t<205\protect \), see Fig. \ref{Figreldist}). Plot d, relative
position of the ghost with respect to the tracer during the first part of the
jet when tracer is sticking to vortex 1. \label{Fig4vortexfastjet}}
\end{figure}

The analysis of the portion of a detected single fast jet is illustrated in
Figs. \ref{Figreldist} and \ref{Fig4vortexfastjet}. This jet corresponds to
a trapping time of the ghosts \( \Delta t\approx 560 \). Since we suspected
that a jet would be located within the sticky zones surrounding the vortex cores,
we plotted in Fig. \ref{Figreldist} (a), (b), and (c) the distances between
the tracer and the vortices versus time, and it is rather clear that the tracer
is always in the vicinity of a vortex during the jet, moreover we also notice
that during its evolution the tracer jumps from one vortex to another: first
the tracer was close to vortex 1, then it jumps to the vortex 3, then to vortex
4 and back to vortex 3. This results is consistent with the observations made
in \cite{Laforgia01} that the sticky zone is the reunion of the surrounding
of all vortices. The possibility of the formation of a pair of two bound vortices
as well as the role of these pairings allowing tracers to jump between vortices
as well as trapping (freeing) them within (from) sticking zones described in
\cite{Laforgia01}, lead us to assume that a pairing was the cause of the odd
behavior of the tracer. This origin is confirmed in Fig \ref{Figreldist} (d),
the distance between the two concerned vortices is plotted and a pairing of
the two vortices is observed for the time interval \( 150<t<200 \). 

In Fig. \ref{Fig4vortexfastjet}, the actual history of the jets is plotted
in various reference frames. In the first figure \ref{Fig4vortexfastjet} (a),
we just plotted the absolute position of the tracer while it evolves in the
jet, which clearly exhibits the difficulty of seeing any order. In the second
figure \ref{Fig4vortexfastjet} (b) the position of the tracer is plotted in
the reference frame which is moving with vortex 3 (The one with whom the tracers
spends most of its time as seen in Fig. \ref{Figreldist} (a)), and the sticking
behavior of the tracer during the jets becomes more clear. In \ref{Fig4vortexfastjet}
(c), we plotted the position of the tracer during the pairing of the vortices
3-4 observed in Fig \ref{Figreldist} (d) in the reference frame whose origin
is the center of vorticity of the pair and which rotates such that the vortices
are stuck oscillating in the x-direction. The double jump from one vortex back
and forth and the exchange between cores is hence illustrated. 
\begin{figure}[!h]
{\par\centering \resizebox*{7cm}{!}{\includegraphics{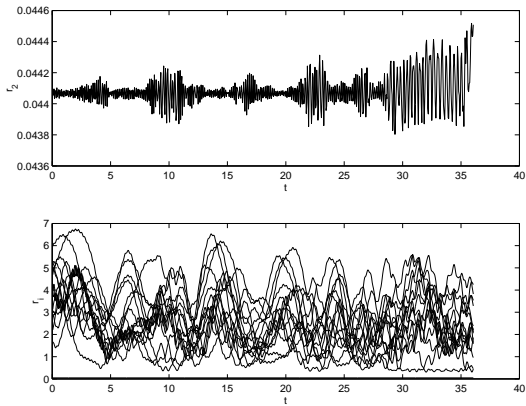}} \par}

\caption{Distance between the tracer and vortices for a coherent jets detected in the
system of 16 vortices. In the upper plot the distance relative to vortex \protect\( 2\protect \)
is plotted, while on the bottom there are the relative distances to all vortices.
We notice towards the end of the jet that while the fluctuations of \protect\( r_{2}\protect \)
increase a pairing occurs between vortex \protect\( 2\protect \) and another
one. This is illustrated by the presence of another vortex nearby the tracer
at the end of the jet.\label{Figjetdetect16v}}
\end{figure}

These visualization of the location of the jets have confirmed the results already
illustrated in \cite{Laforgia01} that the boundaries of the core exhibit the
stickiness. However we would like to emphasize on the fact that in the present
case, this property of the system has been diagnosed with the use of coherent
jets, in other word by analyzing the relative evolution of two nearby trajectories
within a specific scale (phase space ball). In this sense the method used is
rather general, while in \cite{Laforgia01} a more detailed knowledge of the
system was necessary to capture its ``hidden order''. To test even further
the method, we decided to apply it to the system of 16 vortices considered in
section III. Given the previously obtained threshold for 4 vortices, we skipped
the analysis of the distributions of \( \sigma _{L},\: \sigma _{D} \), and
used the similar values to attempt the detection of a fast jet in the flow generated
by 16 vortices. A jet is effectively detected and is illustrated in Fig. \ref{Figjetdetect16v},
meaning that the method is relatively robust since in this 16 vortex system
the cores are much smaller. In this last figure we can notice that a pairing
between two vortices is the cause of the end of the jet, but this may not be
the real end of the jet; both ghosts have escaped but they may have jumped on
the other core, while the tracer is still sticking or vice versa, hence we shall
propose in a following paragraph some possible enhancements to the currently
employed method. Besides this, we also get from Fig. \ref{Figjetdetect16v}
the actual size of the core , which is typically \( r\sim 0.044 \) and is within
a reasonable range from the estimate given by half the minimum distance reached
between vortices \( \min (r_{ij})/2\sim 0.06 \), a fact which was also observed
for 3 and 4-vortex systems \cite{LKZpreprint,Laforgia01}.

In the next section we will consider a little further the notion of the jets
as coherent structure.

\subsection{Jets structure analysis}

In the previous section we were able to characterize that a given tracer was
evolving within what we called a jet, once a given threshold \( \sigma _{D}^{*} \)
for the measured \( \sigma _{D} \) was reached, and then able to visualize
the jet by recording the tracer's position.
\begin{figure}[!h]
{\par\centering \resizebox*{6cm}{!}{\includegraphics{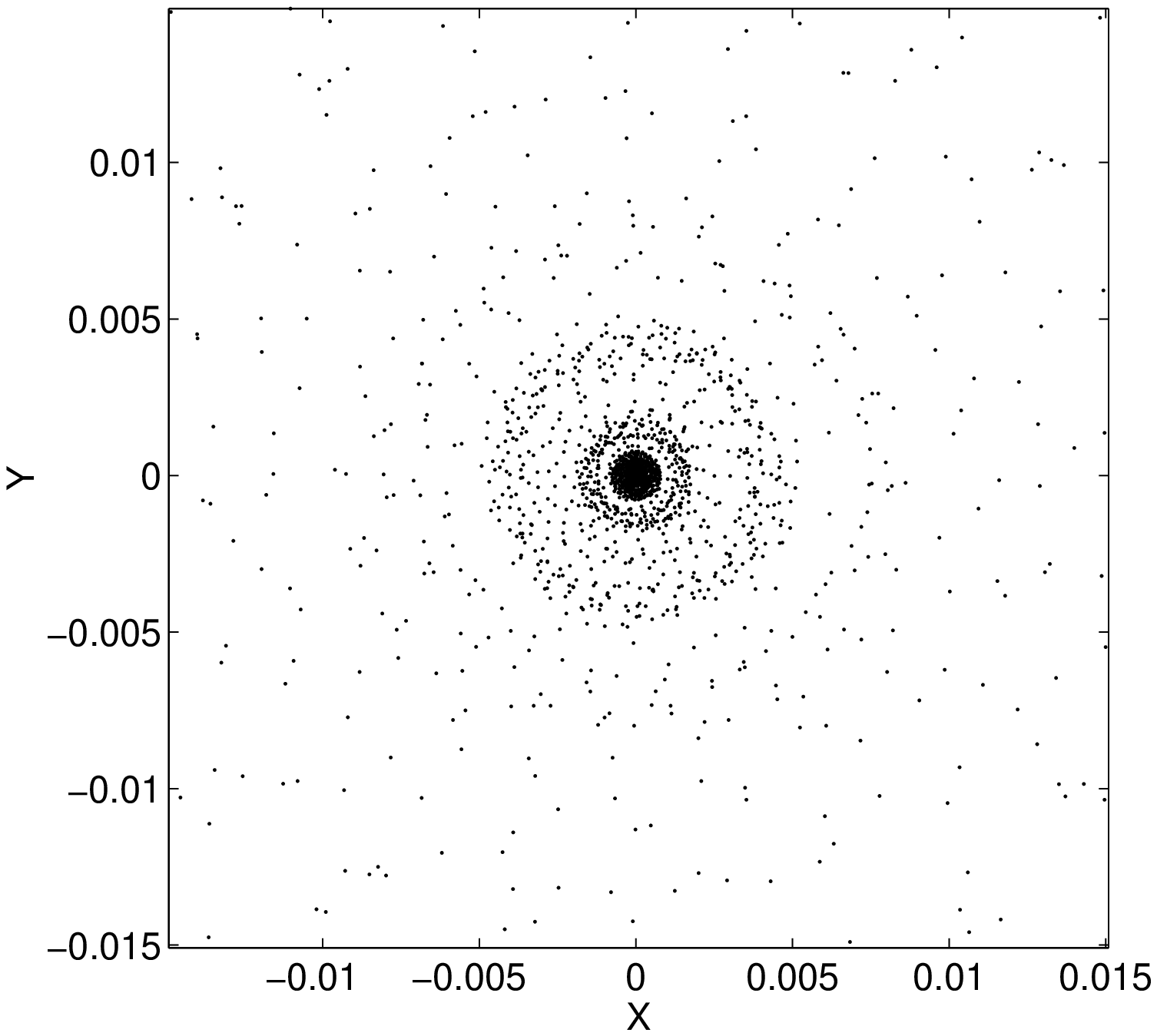}} \\
\par}

{\par\centering \resizebox*{6cm}{!}{\includegraphics{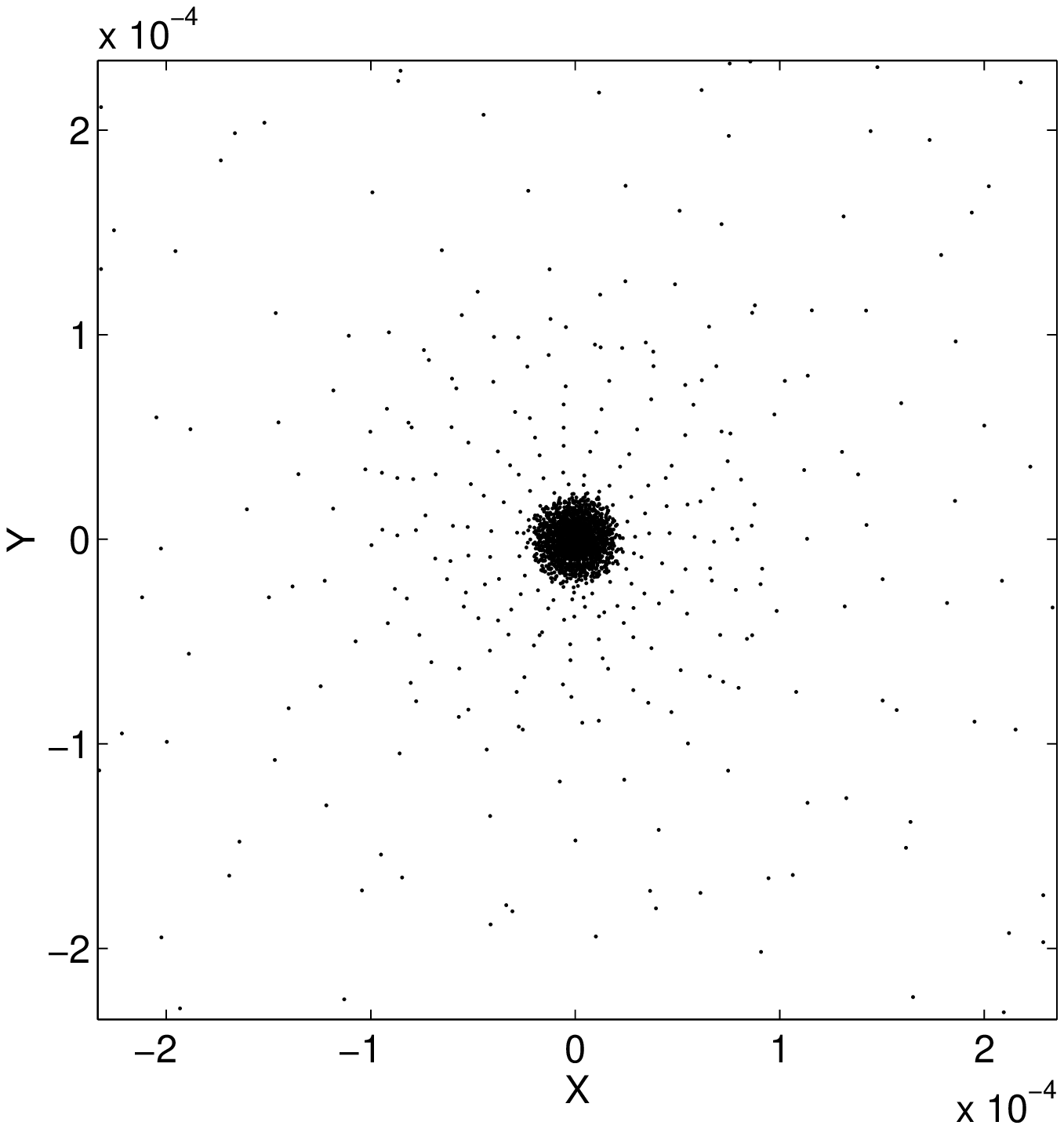}} \par}

\caption{Relative evolution of a ghost within a long lived jets located in the far field
region of the flow generated by four vortices. The lower plot is a centered
zoom of the upper one (see the scales). The distribution within the jets is
clearly not uniform suggesting a possible order organized as ``matroshkas''
(a nested set of jets with increasing radii). \label{Figslowjet}}
\end{figure}
 In the meantime, why not also record the positions of the ghosts, we should
be then able to gather some information about the inner structure of the jet
while practically it should incur a very little numerical overload besides maybe
the need for more storage space.

A first plot of the evolution of a relative ghost position with respect to the
tracer's position is plotted in Fig \ref{Fig4vortexfastjet} (d), where we only
considered the first portion of the jet, when the tracer is sticking to vortex
1. One can see some structure or skeleton within the ball. In order to confirm
this possibility we looked for a longer lived jet, which for the 4 vortex case
was easier to find in the far field region. Plots of such a long lived jet are
presented in Fig. \ref{Figslowjet}, where we effectively see a fine structure
within the jet, which seems to be formed of a hierarchy of circular (tubular)
jets within jets. 
\begin{figure}[!h]
{\par\centering \resizebox*{6cm}{!}{\includegraphics{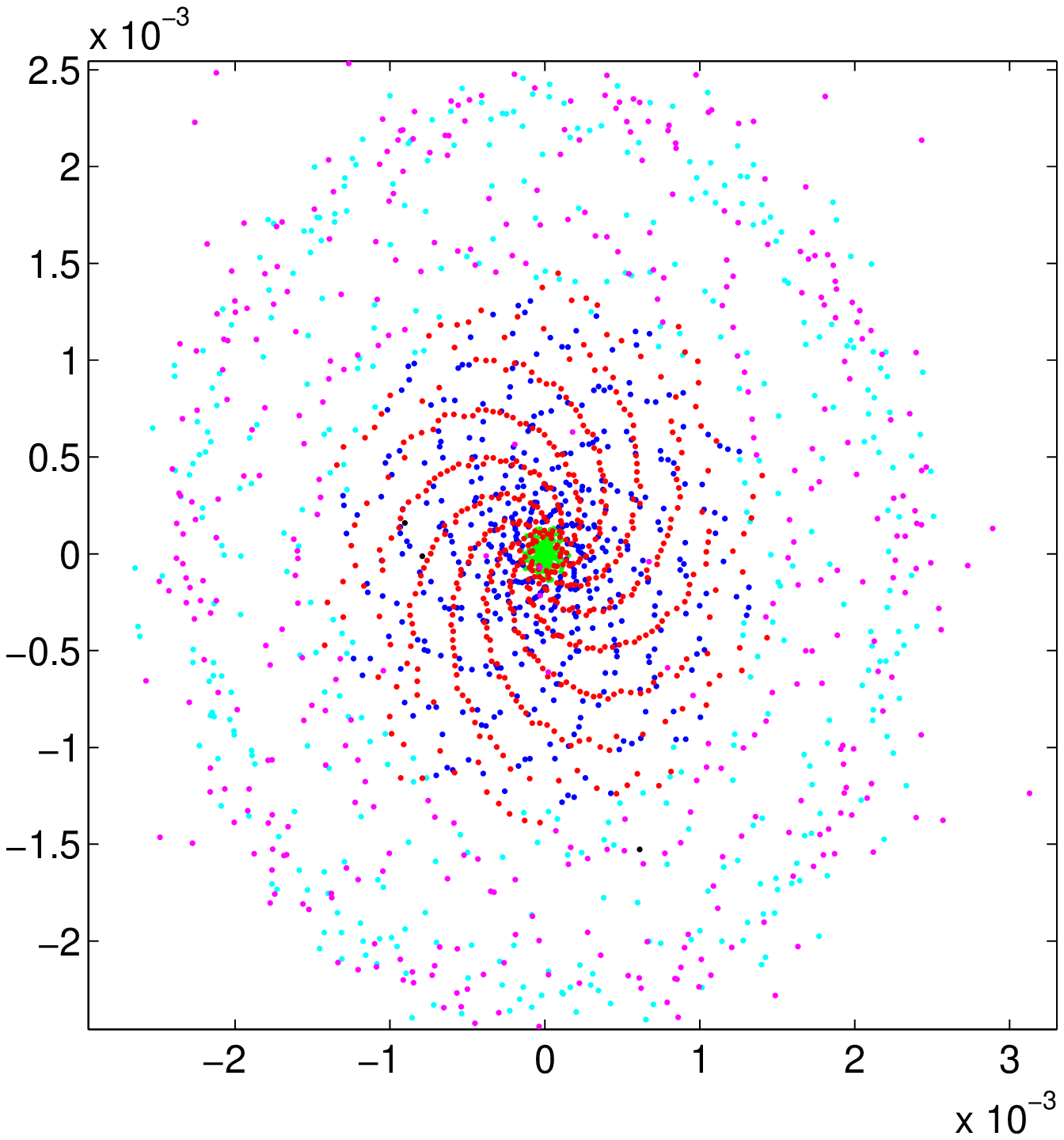}} \\
\par}

{\par\centering \resizebox*{6cm}{!}{\includegraphics{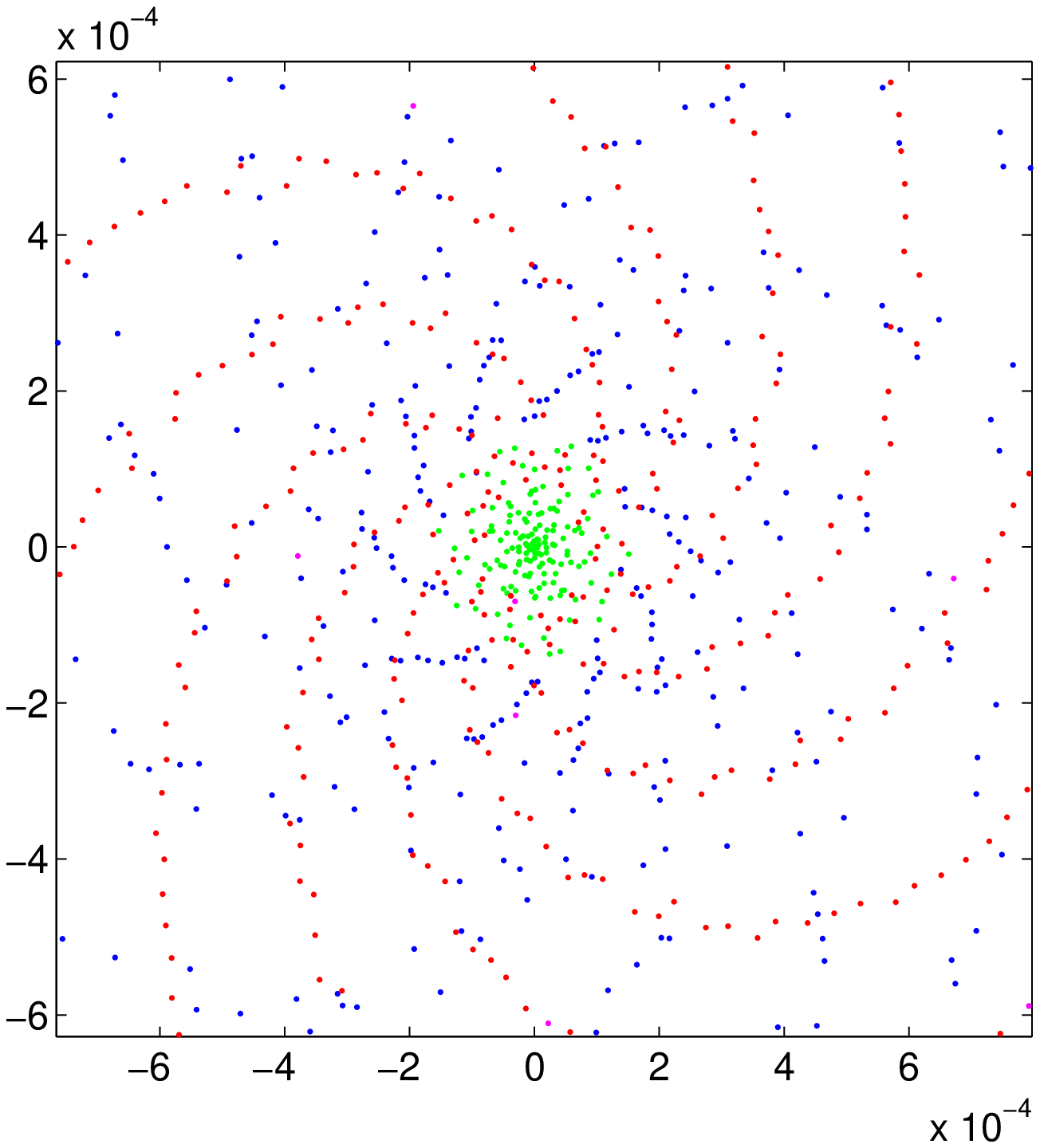}} \par}

\caption{Jet structure for a long lived jet located in the region of strong chaos for
a system of \protect\( 16\protect \) vortices. The lower figure is a zoom of
the upper one corresponding to a magnification of an order of magnitude. The
colors are characterizing different moments of the life of the jet corresponding
to approximatively equal time intervals. They chronogically range as cyan<blue<green<red<magenta.
We see a similar structure of jets within jets as observed in Fig. \ref{Figslowjet},
and ghosts spiraling back and forth between them.\label{Fig16vjettime}}
\end{figure}
 Wondering if these features were general or specific to these 4 vortex systems,
we decided to consider the fast jet illustrated in Fig.\ref{Figjetdetect16v}
and check its structure. The results are plotted in Fig. \ref{Fig16vjettime},
where the relative position of the ghost is colored differently for different
time periods of the life of the jets. We can see that effectively the nested
set of jets within jets remains, and that the ghosts is also spiraling back
and forth in between. This figure is also informative in the sense that we actually
see the ghost going back very close to the tracer. In other words, the area
characterized by the green points in Fig. \ref{Fig16vjettime} does not correspond
to the beginning of the jet and therefore is not an artifact of having initially
placed the ghost in the vicinity of the tracer. 

We shall close this subsection with a final remark on this ``matroshka'' structure
of the jets. Namely, this nested structure suggests that for each identified
jet we can define a suite \( (r_{n}) \), such that \( r_{n} \) is a decreasing
function of \( n \) with \( r_{n}\rightarrow 0 \) and \( n\rightarrow \infty  \)
and \( r_{0} \) is for instance determined by the largest tube (jet) seen in
Fig. \ref{Figslowjet} or Fig. \ref{Fig16vjettime}. To each subjet \( n \)
we can assign a distribution of trapping time \( \rho _{n}(\tau ) \) as well
as a transit time \( t_{n} \) (or a distribution) associated with the tracer
spiraling from one subjet \( n \) to one of its two neighbors \( n-1 \) or
\( n+1 \). In this light depending on the transport properties of the whole
system are likely to depend of the distributions \( \rho _{n}(\tau ) \) and
the ratio \( r_{n}/r_{n+1} \), and for instance if we have in the limit \( n\rightarrow \infty  \)
both \( r_{n}/r_{n+1}\rightarrow r_{\infty } \) and \( \rho _{n}(\tau )\rightarrow \rho _{\infty }(\tau ) \),
the FSLE become effectively independent of scale within the jet. This hierarchical
structure is also reminiscent of the discrete renormalization group, and hence
we can speculate that log-periodic oscillation described in \cite{Benkadda99}
may be observed.

\section{Transport Properties}

\subsection{Definitions }

For the considered case, all vortices have positive strength and therefore are
moving within a finite domain. It is important to define what quantities will
be measured to characterize the transport properties of the system. There has
been evidence in \cite{Boatto99} of radial diffusion, but the diffusion coefficient
\( D \) is vanishing as \( R\rightarrow \infty  \) with typical behavior \( D\sim 1/R^{6} \).
In the case of more than four vortices we can still expect a similar type of
behavior. However the far field region is of little interest when one want to
address the transport properties of typical geophysical flows since the most
relevant is the region being accessible to the vortices, \emph{i.e} also called
the region of ``strong chaos''. In fact the same problem was faced when considering
system of three point vortices. For these particular systems the vortex motion
is integrable, which has for consequence to restrict the accessible domain of
tracers within a finite region surrounded by a KAM curve. The way around this
problem was then to measure how many times a tracer rotates around the origin,
in other words measure the diffusion along an angular direction. However even
though suitable, this solution may face some criticisms due to the singularity
present at the origin. In this article we consider tracer transport in the way
already used with some success in \cite{LKZpreprint}. Namely the arclength
\( s(t) \) of the path traveled by an individual tracer up to a time \( t \),
which writes
\begin{equation}
\label{arclenghtdefinition}
s_{i}(t)=\int ^{t}_{0}v_{i}(t')dt'\: ,
\end{equation}
where \( v_{i}(t') \) is the absolute speed of the particle \( i \) at time
\( t' \). One advantage of this quantity is that it is independent of the coordinate
system and as such we can expect to infer intrinsic properties of the dynamics.
The main observable characteristics will be moments of the distance \( s(t) \)
defined in (\ref{arclenghtdefinition}):
\begin{equation}
\label{momentsdefi}
M_{q}=\langle |s_{i}(t)-\langle s_{i}(t)\rangle |^{q}\rangle \: ,
\end{equation}
where \( i \) corresponds either to the i-th vortex or a tracer in the field
of 4 or 16 vortices. The averaging operator \( \langle \cdots \rangle  \) needs
a special comment. Expecting anomalous transport one should be ready to have
infinite moments starting from \( q\ge q_{0} \). Particularly it can be \( q_{0}=2 \).
The value of \( q \) will vary from \( 0 \) to \( 8 \). To avoid any difficulty
with infinite moments we consider truncated distribution function 
\begin{equation}
\label{truncdist}
\rho _{tr}\left( s_{i}\right) =0\: ,\hspace {10mm}s_{i}>s^{*}_{i}\: .
\end{equation}
The condition (\ref{truncdist}) was discussed in details in \cite{Weitzner01}
to satisfy the physical restriction of a finite velocity. This condition put
some constraints on the maximum value \( q* \) of \( q \) corresponding to
the maximum time \( t* \), beyond which the moments are basically monitoring
the population of almost ballistic trajectories.

Up to the mentioned constraints, we will always consider the operation of averaging
to be performed over truncated distributions. In this perspective all moments
are finite and one can expect
\begin{equation}
\label{momentexpectation}
M_{q}=\langle |s_{i}(t)-\langle s_{i}(t)\rangle |^{q}\rangle \sim D_{q}t^{\mu (q)}\:,
\end{equation}
with, generally, \( \mu (q)\ne q/2 \) as one would expect from normal diffusion.
The nonlinear dependence of \( \mu (q) \) means multifractality of the transport.
Some authors use the notion of weak (\( \mu (q)=const\cdot q \)) and strong
(\( \mu (q)\ne const\cdot q \)) anomalous diffusion \cite{Castiglione99,Andersen00}
or strong and weak self-similarity \cite{Ferrari01}. For more information on
the appearance of multi-fractal kinetics and related transport see \cite{Zaslav2000,Zaslav2000_1}.

\subsection{Vortex transport properties in 16 vortex system}

\begin{figure}[!h]
{\par\centering \resizebox*{7cm}{!}{\includegraphics{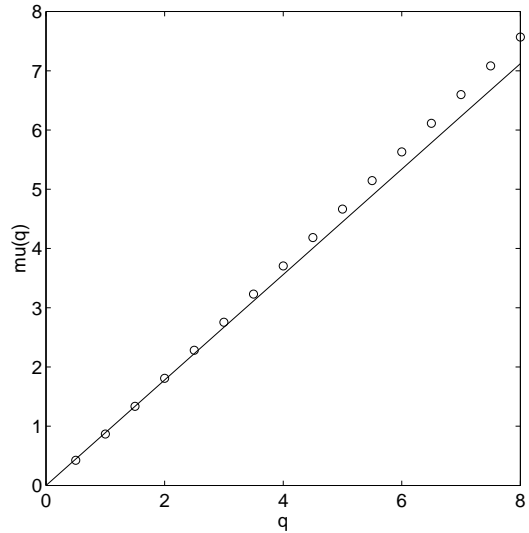}} \par}

\caption{Large-time behavior for the different moments \protect\( \langle |s(t)-<s(t)>|^{q}>\sim t^{\mu (q)}\protect \)
for the flow of 16 point vortices, with \protect\( \mu (2)\approx 1.80\protect \)
corresponding to a superdiffusive regime.\label{Figtransport16v}}
\end{figure}

Due to the low dimensionality of a system of 4 point vortices, no typical transport
behavior has been measured, but there is an analysis of the phase space topology
and Poincare sections \cite{Laforgia01}. The results for 16 vortices were computed
by using a number of different ``equivalent'' initial conditions. Such different
trajectories were obtained by letting a given initial condition evolve and record
the positions of the vortices at different times with a given accuracy (typically
\( 10^{-5} \)), while making sure no vortices were initially involved in pairings.
The strong chaoticity of the system leads very rapidly to different trajectories
while this choice of initial conditions let us keep the constant of motions
within a relative small error. The transport properties are then obtained by
averaging over all vortices as well as over different sets of trajectories that
correspond to our specific arbitrary choice of the constants of motion. The
results are presented in Fig. \ref{Figtransport16v} 
, where the exponent characterizing the long time power law behavior of different
moments is plotted versus moment order. The actual value of the second moment
which is typically used to characterize the transport properties, is measured
to be \( \mu (2)\approx 1.80 \) and indicates a strong superdiffusive behavior
of the vortex subsystem. This behavior can be linked to the pairing of vortices
which induces a strong quasi-ballistic dynamics of both vortices of the pair.
Indeed let us compare the exponent \( \gamma \approx 2.68 \) characterizing
the distribution of time fro which vortices are trapped into a pair measured
in Fig. \ref{Figpariingtimedist}, and the value corresponding to the second
moment behavior from Fig. \ref{Figsnapshot16v}. We see a good agreement with
the expected law \( \gamma \approx \mu (2)-1 \) (see Table \ref{Table1}) \cite{KZ2000,Zaslavsky2000}.
From this observation we can conclude that the superdiffusive behavior of a
vortex from the system of 16 vortices is related to the pairing phenomenon. 
\begin{table}[!h]
{\centering \begin{tabular}{|c|c|c|}
\hline 
&
16 vortices&
tracers in 4 vortex flow\\
\hline 
\hline 
\( \mu  \)&
\( 1.80 \)&
\( 1.82 \)\\
\hline 
\( \gamma  \)&
\( 2.68 \)&
\( 2.82 \)\\
\hline 
\end{tabular}\par}

\caption{Trapping times and second moment behavior, for 16 vortices the trapping time
is identified with pairing times, while for the tracers the trapping times corresponds
to the time spent within a jet (i.e sticky zone). We can notice the very good
agreement with the estimate given by \protect\( \gamma =1+\mu \protect \).\label{Table1}}
\end{table}
 We now shall investigates the properties of tracers and first start with the
chaotic flow generated by four vortices.

\subsection{Passive tracers in 4-vortex system}

\begin{figure}[!h]
{\par\centering \resizebox*{7cm}{!}{\includegraphics{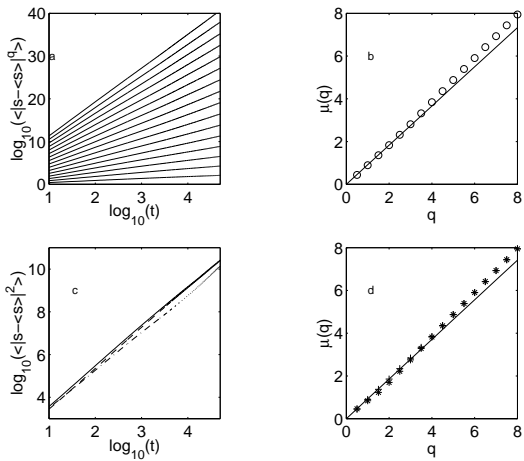}} \par}

\caption{Plot of different transport properties of tracers in 4-vortex system. Plot
a, the different moment versus time are plotted for \protect\( q=1/2,1,\cdots ,8\protect \).
Plot b, long time exponent behavior for the different moments (see (\ref{momentsdefi})),
with \protect\( \mu (2)\approx 1.83\protect \) corresponding to a superdiffusive
regime. Plot c, second moment versus time, the full line corresponds to the
whole data, the dashed line is computed with all trajectories but the parts
sticking to vortices, and the dot-dashed line to all trajectories but the part
sticking to the outer region. Plot d, different exponents for partial data,
{*} outer region cut (\protect\( \mu _{f}=1.7\protect \)), + vortex-sticking
cut (\protect\( \mu _{s}=1.85\protect \)); in both cases the transport is anomalous
and superdiffusive.\label{Figtransport4vtrac}}
\end{figure}

The results describing the transport properties of passive tracers are illustrated
in Fig. \ref{Figtransport4vtrac}. We also observe strong anomalous behavior
for the tracers with an exponent for the second moment \( \mu (2)\approx 1.82 \).
This behavior was previously observed in \cite{KZ2000} for integrable flows
driven by three vortices. In this latter case, the origin of the anomalous behavior
was directly linked to the presence of islands of regular motion within the
stochastic sea and the phenomenon of stickiness observed around them. In the
present case the driving flow is chaotic but nevertheless the existence of cores
surrounding the vortices and the far field region allows a direct analogy and
we may say that the tracer meet similar structures. The only difference is that
some of the structure elements (the cores) are ``mobile''. Moreover the previous
study \cite{KZ2000,Laforgia01} shows that tracers effectively stick to the
vortex cores, and the FTLE field presented in \cite{Boatto99} shows that in
the far field region the motion of tracers is quasi ballistic, which is a good
indication of a sticking behavior.

This conclusion can be strengthened by considering previously described jets.
We have seen that the jets are located in the sticky regions, and are either
moving around cores and jumping from core to core or rotating in the far field
region. Since we now have ``mobile'' sticky regions traced by the long lived
jets which we are able to monitor, we can estimate that the sticking time (or
trapping time in the sticky zone) is more or less similar to the time it takes
for a ghost to escape a long lived jet, hence the power law behavior observed
in Fig. \ref{Figescapetime4vortex} characterize the exponent \( \gamma  \)
for trapping time. We then actually observe an exceptional agreement with the
expected \( \gamma \approx \mu (2)-1 \) relation (see Table \ref{Table1}).
Hence since the notion of jet as defined in Section IV is relatively general
when compared to the notion of core or far field region, we shall say that the
anomalous diffusion finds its origin in the existence of long lived ``coherent''
jets of passive tracers for the chaotic flow generated by four vortices. Since
we know that two different types of jets exists (see Fig. \ref{FIgspeeddist})
and are able to differentiate them easily, in a similar manner as for the three-vortex
case \cite{KZ2000,LKZpreprint} we are able to quantify the influence of each
type of jet on the transport properties of the system by discarding tracers
evolving within a given type of jets. The results of this analysis are presented
in Fig. \ref{Figtransport4vtrac}, plots c and d. We can conclude that both
type of jets are giving rise to anomalous transport with very close characteristic
exponents and that they both contribute to the observed strong anomalous diffusion.

\subsection{Passive tracers in 16-vortex system}

\begin{figure}[!h]
{\par\centering \resizebox*{7cm}{!}{\includegraphics{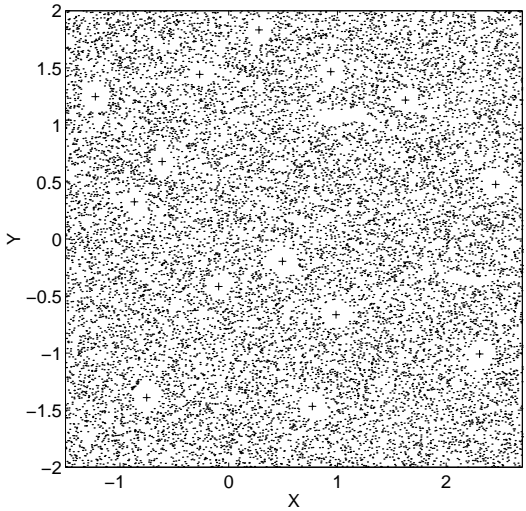}} \par}

\caption{Local snapshot of the system of \protect\( 16\protect \) vortices with \protect\( 9.\: 10^{4}\protect \)
tracers. The vortices are located with the ``+'' sign. We can see the cores
surrounding the vortices. The snapshot is taken early in the simulation to prevent
too much dispersion of the tracers and visualize the core. As a consequence
we can expect that the actual size of the cores to be a little smaller than
the radius of \protect\( r\approx 0.1\protect \) one can measure on the snapshot.\label{Figsnapshot16v}}
\end{figure}

\begin{figure}[!h]
{\par\centering \resizebox*{7cm}{!}{\includegraphics{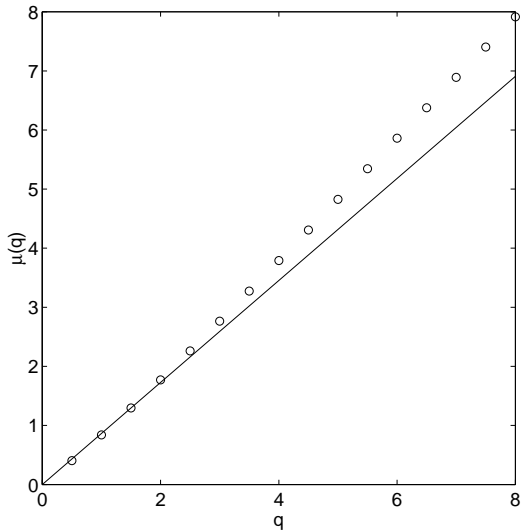}} \par}

\caption{Exponents \protect\( \mu (q)\protect \) for the tracer moments for the flow
driven by 16 point vortices, with \protect\( \mu (2)\approx 1.77\protect \).The
slope for small \protect\( q\protect \) is \protect\( 1.75\protect \) and
for large \protect\( q\protect \) is \protect\( 1.95\protect \).\label{Figtransptracer16v}}
\end{figure}

In both previous cases we observed a signature of anomalous super diffusion
with \( \mu (2)\approx 1.8 \). Let us now investigate the transport properties
of passive tracers in a flow generated by 16 identical point vortices, described
partially in Section III. there are long lived jets within the region of strong
chaos (see Fig. \ref{Fig16vjettime}) for the flow and the detected jet lives
in a vicinity of a vortex and is probably sticking to its core. A snapshot of
the system at an early stage is given in Fig. \ref{Figsnapshot16v} which effectively
identifies the cores surrounding the vortices.
 The results describing the transport properties of passive tracers are in Fig.
\ref{Figtransptracer16v} and show strong anomalous behavior with the exponent
\( \mu (2)=1.77 \). 

Eventhough we only consider one arbitrary initial condition of the vortex system,
it is reasonable to assume that the transport properties obtained for such a
system are fairly general in the sense that they persist with an increase of
the number of vortices if the probability of pairing persists. That means a
possibility to neglect the occurrence of more complicated long lived clustering
such as the triplet observed in Fig. \ref{FIr16vpairingandtriple}, quadruples,
etc...

\section{Conclusion and Discussions}

In this paper we have investigated the dynamical and statistical properties
of the vortex and passive particle advection in chaotic flows generated by four
and sixteen point vortices. The goal of the work was to provide qualitative
insights on general transport properties of two-dimensional flows, especially
geophysical ones, imposed by the topology of the phase space. The system of
16 vortices can be considered as a fairly large system while the 4-vortex one
is the minimal one with chaotic dynamics of the vortices. The time-averaged
spatial distribution of the vortices is characterized by a non-uniform density
of vorticity, which implies strong vortex-vortex correlation. These correlations
manifest themselves by a phenomenon of stickiness, namely the formation of long
lived pairs of vortices, triplets, etc... Since both these structures are integrable,
we can speculate that the stickiness occurs by forming quasi-integrable subsets.
The clusters of vortices can form a hierarchical or nested set of integrable
subset and hence put some constraints on the frame of possible larger structures
involving more vortices. The statistics of pairing time exhibits a power law
tail confirming in that way the sticky nature of the phenomenon, which in return
allows us to make an analogy with an ideal dynamical system and give a good
analytical estimate of the characteristic exponent of the pairing times distribution.

The chaotic nature of the governing flows did not allow the use of diagnostics
such as Poincar\`e maps or distribution of recurrences commonly used for systems
with \( 1\: 1/2 \) degrees of freedom, hence a new technique inspired by Finite
Time Lyapunov Exponents diagnostics has been put into place. Passive tracer
motion is analyzed by measuring the mutual relative evolution of two nearby
tracers. A possibility of tracers to travel in each other vicinity for relatively
large times, confirmed the presence of a hidden order for the tracers which
we call jets. The jets can be understood as moving clusters of particles within
a specific domain where the motion is almost regular from a coarse grained perspective.
The chaotic nature of the motion is confined within the characteristic scale
of a given jet, where nearby tracers are trapped. The distribution of trapping
times in the jets shows a power-law tail whose characteristic exponent is quantitatively
very similar to the one related to pairing times. The calculated Lyapunov exponent
for tracers within the jets exponents are similar to what is called Finite Sized
Lyapunov Exponent \cite{Ferrari01}, but in our calculations no average is taken
and thus a distribution is obtained. These distributions of the Lyapunov exponents
exhibit a finite local minimum, which results from the competition of the trapping
in jets and the strongly chaotic motion outside the jets. The existence of this
minimum allows a dynamical test which identifies if a tracer is trapped within
a jet and thus to localize the jet in phase space. Jets are found to exist on
the boundary of the cores surrounding the vortices as well as on the outer rim,
the region to which vortices have no access to. This behavior is analogous to
the sticking one observed in the three vortex case, and thus can be incorporated
to the general notion of the phenomenon of stickiness. The differentiation between
the two possible types of jets is completely determined by the ratio of the
corresponding pair of Lyapunov exponents. We thus obtain de facto a diagnostic
to locate coherent structures. The method is then successfully applied to the
system of sixteen vortices, leading to a possible general dynamical mechanism
of detecting coherent structures. Further analysis of the structure of the jets
itself reveals a complicate nested structure of jets within jets, which indicates
that jets exists at different scales. The distribution of trapping times within
jets is computed and shows a power-law tail whose characteristic exponent is
similar to one observed for vortex pairing time. Since the trapping in jets
is also a stickiness phenomenon we may assume that the analytical estimate given
for the vortex pairing time is also valid for the trapping time within jets.
The transport properties of the 16 vortices as well as those of the tracers
in both systems of 4 and 16 vortices are found to be anomalous with characteristic
exponent \( \mu \sim 1.75-1.8 \). All these results are in good agreement with
the trapping times characteristic exponents and the kinetic theory presented
in \cite{KZ2000,LKZpreprint}. Moreover the transport properties are all of
the multi-fractal type or strongly anomalous in the sense defined in \cite{Castiglione99,Andersen00}.
This property of the transport is a consequence of the existence of different
sticky zones and the related structures in phase space.

\acknowledgments{The authors would like to express their thanks to L. Kuznetsov for very useful discussions. This work was supported by the U.S. Navy Grants Nos. N00014-96-1-0055 and N00014-97-1-0426, and the U.S. Department of Energy Grant No. DE-FG02-92ER54184.}

\end{document}